\documentclass[fleqn,10pt]{wlscirep}
\usepackage[utf8]{inputenc}
\usepackage[T1]{fontenc}
 \pdfoutput=1
\usepackage{subfigure}
\usepackage{multirow}
\usepackage{bm}
\title{Solid state phase transformation kinetics in Zr-base alloys}

\author{A. R. Massih}
\author{Lars O. Jernkvist}
\affil{Quantum Technologies AB, Uppsala Science Park, Uppsala SE-75183, Sweden}

\begin{abstract}
We present a kinetic model for solid state phase transformation ($\alpha \rightleftharpoons \beta$) of common zirconium alloys used as fuel cladding material in light water reactors. The model computes the relative amounts of $\beta$ or $\alpha$ phase fraction as a function of time or temperature in the alloys. The model accounts for the influence of excess oxygen (due to oxidation) and hydrogen concentration (due to hydrogen pickup) on phase transformation kinetics.  Two variants of the model denoted by A and B  are presented. Model A is  suitable for simulation of laboratory experiments in which the heating/cooling rate is constant and is  prescribed. Model B is more generic. We compare the results of our model computations, for both A and B variants, with accessible experimental data reported in the literature covering heating/cooling rates of up to 100 K/s. The results of our comparison are satisfactory, especially for model A. Our model B is intended for implementation in fuel rod behavior computer programs, applicable to a reactor accident situation, in which the Zr-based fuel cladding may go through $\alpha \rightleftharpoons \beta$ phase transformation.
\end{abstract}
\begin{document}

\flushbottom
\maketitle
%
%
\thispagestyle{empty}

\section{Introduction}
\label{sec:intro}
Alloys based on group IVB metals of the periodic table, comprising titanium (Ti), zirconium (Zr) and hafnium (Hf), are widely utilized in aerospace and marine, nuclear energy, and biomedical products \cite{Lutjering_Williams_2007,Boyer_2010,zhang2019review,Lemaignan_Motta_1994,suzuki2020appraising,Tricot_1992}. They share common chemical, physical and electronic attributes; in contrast to disparate nuclear properties\cite{Tricot_1992}, e.g. Zr versus Hf. In particular, in the solid state they are in two stable crystal structures: low temperature $\alpha$ (hexagonal closed-packed) and high temperature $\beta$ (body-centered cubic).

Zirconium base alloys are materials of choice  for fuel cladding  in light water reactors because of their low neutron absorption cross section and good corrosion resistance. The alloys comprise Zircaloy-2 (1.5wt\%Sn, 0.12wt\%Fe, 0.05wt\%Ni, 0.1wt\%Cr, 0.12wt\%O) for boiling water reactor,  and Zircaloy-4 (1.5wt\%Sn, 0.15wt\%Fe, 0.1wt\%Cr,0.12wt\%O) or Zr1wt\%Nb0.1wt\%O (or Zr1NbO) among others for pressurized water reactor.

Zirconium alloys in solid state undergo a phase transformation from  $\alpha$ phase (space group P$6_3$/mmc) to $\beta$ phase (space group IM3m) at ambient pressures  \cite{Lemaignan_Motta_1994}.  Solid state phase equilibria of these alloys basically comprise three phase domains, namely, the $\alpha$ phase, ($\alpha+\beta$) coexisting phase, and the $\beta$ phase. The phase equilibria have been evaluated for Zircaloy-4 in \cite{Miquet_et_al_1982a} and for the Zr-Nb-O system in \cite{Perez_Massih_2007}. The starting temperature of the $\alpha \to (\alpha+\beta)$ transition for Zircaloy-4 is about 1090 K and for the Zr1NbO alloy is around 1040 K \cite{Kaddour_et_al_2004}. Similarly, the start of the $(\alpha+\beta) \to \beta$ transition is about 1250 K for Zircaloy-4 and 1210 K for Zr1NbO \cite{Kaddour_et_al_2004}. The phase transition is first order, meaning that  a latent heat is produced  during the transition, amounting to $\Delta H \approx 4$ kJmol$^{-1}$ \cite{Terai_et_al_1997}.

The solid state phase transformation kinetics under nonisothermal conditions for Zr-alloys has been modeled by means of nonspatial dependent  (point kinetics) differential equation method previously \cite{Forgeron_et_al_2000,Massih_Jernkvist_2009,Massih_2009}. The method described in \cite{Massih_Jernkvist_2009,Massih_2009} posited that the rate of change for  the volume fraction of the transformed phase is a functional of the fraction of the transformed phase in thermal equilibrium  and the relaxation time for the phase transition which are both functions of temperature. It fitted the available  experimental  data on Zircaloy-4 and Zr1NbO well for constant heating/cooling rates of up to 100 K/s. However, parameters in this model, henceforth referred to as model A, depend explicitly on the heating/cooling rate $q = dT/dt$, and its applicability is limited to cases where $q$ is constant and known in advance. Model B, introduced in the present paper, is free from this constraint.

The rest of the paper is organized as follows. In section \ref{sec:expdata}, we briefly review experimental data in the literature pertinent to our modeling. In section \ref{sec:models}, we describe the aforementioned model A and model B plus the modified parameters that account for excess oxygen and hydrogen concentration in the material. In section \ref{sec:compute}, we compare our model computation results with the experimental data surveyed in section \ref{sec:expdata}. In section \ref{sec:discuss}, we analyze our model and discuss its results and its applicability. A summary and conclusions are presented in section \ref{sec:conclude}.

\section{Experimental data}
\label{sec:expdata}
In this section, we provide a brief survey of published experimental data on solid-to-solid ($\alpha \rightleftharpoons \beta$) phase transformation of zirconium alloys used  widely in the nuclear industry. In particular, we go over the data that we have utilized in our modeling and analysis. The alloys discussed here are listed in table \ref{tab:chemcomp} with their nominal  chemical compositions. We first recount the phase equilibria properties  in regard to the excess oxygen concentration in the alloy and the hydrogen content of the alloy. Then we peruse data on phase transformation of the alloys during time varying temperature, namely continuous heating and cooling.

\begin{table}[ht]
\centering
\begin{tabular}{|l|l|l|l|l|l|}
\hline
Alloy & \multicolumn{5}{c|}{Alloying elements, wt\%}\\
\hline
& Sn & Fe & Cr & Nb & O \\
\hline
Zircaloy-4  & 1.5 & 0.21 & 0.1 & \dots &  0.125\\
Zircaloy-4L & 1.3 & 0.21 & 0.1 & \dots &  0.125\\
Zr1NbO(M5)  & \dots & \dots & \dots  & 1.0 & 0.125\\
E110        & \dots & \dots & \dots  & 1.0 & 0.05\\
ZIRLO       &  0.94 & 0.09 & \dots  & 0.9 & 0.120\\
\hline
\end{tabular}
\caption{\label{tab:chemcomp} Nominal chemical composition of some Zr-base alloys.}
\end{table}

\begin{figure}[!hbtp]
  \begin{center}
    \includegraphics[width=0.7\textwidth]{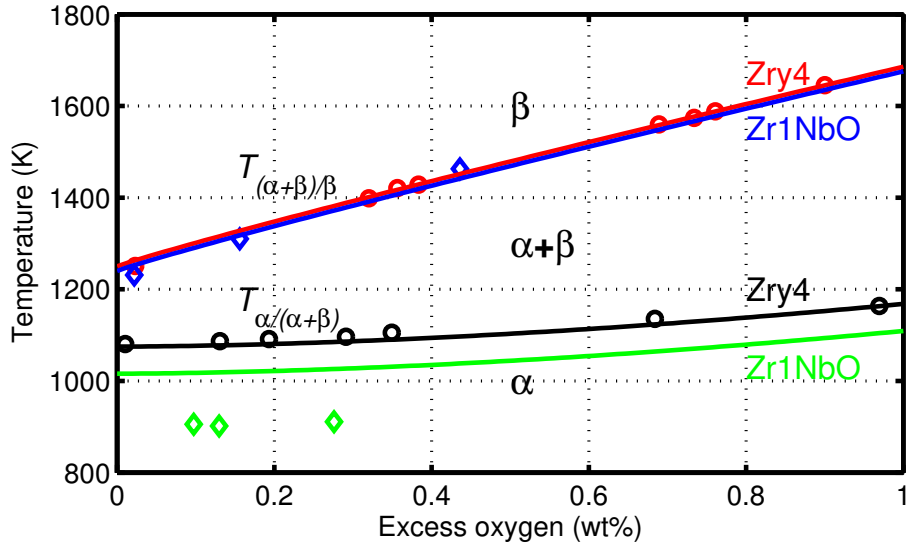}
    \caption{Phase boundary temperatures  versus metal excess oxygen concentration  for Zircaloy-4$\equiv$Zry4 (circles) \cite{Chung_Kassner_1979},
              Zr-0.5\%NbO  (blue diamonds) \cite{Hunt_Niessen_1970}, and Zr-1\%Nb  (green diamonds) \cite{Toffolon_et_al_2002}. The lines are empirical correlations for  Zry4 and Zr1NbO alloy.}
    \label{fig:zry-o}
  \end{center}
\end{figure}

The phase boundary temperature versus excess oxygen concentration for Zircaloy-4 was determined by Chung and Kassner \cite{Chung_Kassner_1979}. These authors studied Zircaloy/oxygen phase diagram by resistivity measurements and by metallographic examination of equilibrated specimens. Chung and Kassner determined the location of phase boundary of oxidized Zircaloy-4  specimens annealed to form uniform oxygen distribution in the $\alpha$ phase region. Similar specimens were brought out to equilibrium temperatures in $\beta$ phase and in ($\alpha+\beta$) mixed phase and then quenched to obtain the corresponding phase boundaries. We have extracted phase boundary data, i.e. $\alpha/(\alpha+\beta)$ and $\beta/(\alpha+\beta)$ transus, determined by metallographic examination from table 2 of \cite{Chung_Kassner_1979}. These data are displayed here in Fig. \ref{fig:zry-o} by circles.

The phase boundary temperature versus excess oxygen concentration for certain ZrNbO alloys for $\beta/(\alpha+\beta)$ transus was determined by Hunt and Niessen \cite{Hunt_Niessen_1970}. The authors investigated this transus for ZrNbO alloys containing up to 5 wt\% niobium by metallographic examination. The oxygenated specimens were heated into the $\beta$ phase region then quenched in water. Thereafter, they examined the microstructure of the samples for retained $\alpha$ phase in the transformed $\beta$ matrix. Their data on Zr0.5wt\%NbO were used to our intent for Zr1NbO, which are shown in figure \ref{fig:zry-o} by blue diamonds. Because oxygen is an alloying element in Zircaloy-4 and  Zr1NbO (table \ref{tab:chemcomp}), to obtain the excess oxygen, we have subtracted 0.12 w\% from the absolute values of the aforementioned data.

A few data points for low excess oxygen concentrations are available for Zr1NbO close to the  $\alpha/(\alpha+\beta)$ boundary \cite{Toffolon_et_al_2002}. These data are shown in Fig. \ref{fig:zry-o} as green diamonds. Toffolon et al. \cite{Toffolon_et_al_2002} determined these data by calorimetry starting from a Zr1wt\%Nb binary alloy. A more detailed assessment of the phase equilibria of the Zr-Nb-O ternary system can be found in \cite{Perez_Massih_2007}.

Kinetics of solid state phase transformation in Zr-based alloys has been investigated experimentally over the years. Early studies include references \cite{Corchia_Righini_1981,Arias_Roberti_1983,Arias_Guerra_1987}, which used electrical resistivity versus temperature measurements to determine the $\alpha/\beta$ phase boundaries in Zircaloy-2. Corchia and Righini studied phase transformation  \cite{Corchia_Righini_1981} in the temperature range of 1000 - 1400 K with heating rates in the range 200 - 1400 K/s and cooling rates 10 - 60 K/s. They concluded that, in  nonequilibrium conditions, the kinetics of the phase transformation is strongly influenced by the microstructure of the specimen and  by inhomogeneous distribution of the fast diffusing elements (Cr, Fe, Ni). Arias and coworkers  employed much slower heating/cooling ($\approx 0.05$ K/s) to establish the $\alpha/\beta$ phase boundaries \cite{Arias_Guerra_1987}. They also evaluated \cite{Arias_Roberti_1983} the influence of tin, iron and oxygen on the ($\alpha+\beta$) field. The aforementioned early studies provide us with some experimental facts and are not in a form that could be used for direct comparison with our modeling.

A systematic study of $\alpha/\beta$  phase transformation of stress-relieved annealed (SRA)  Zircaloy-4 and recrystallized (RX) Zr1NbO alloy was carried out by Forgeron et al. \cite{Forgeron_et_al_2000}. The study includes phase transformation under quasi-equilibrium conditions, i.e. very low heating/cooling  rates, ranging from 0.002 to 0.33 K/s. The researchers used DSC, i.e. Differential Scanning Calorimetry, to determine the fraction of $\beta$ (or $\alpha$) phase as a function of temperature in the alloys. They also made direct observation of microstructure by optical microscopy on samples annealed for a few hours at different temperatures in the $(\alpha+\beta)$ two-phase domain. The experimental uncertainty of the phase fraction measurement was estimated to be around $\pm 0.05$ or less.

In order to study the kinetics of phase transformation,  Forgeron et al.  \cite{Forgeron_et_al_2000} utilized dilatometric technique on small samples (12 mm in length) machined from  as-fabricated cladding tubes. The dilatometric measurements for Zircaloy-4 samples included data with heating/cooling rates of 10 and 100 K/s but only with heating rates 10 and 100 K/s for Zr1NbO alloy. As in their DSC measurements, the uncertainty of the phase fraction measurement is reported to be  $\le0.05$ and $\pm 10$ K for measured temperatures.

\begin{figure}[!hbtp]
  \begin{center}
    \includegraphics[width=0.7\textwidth]{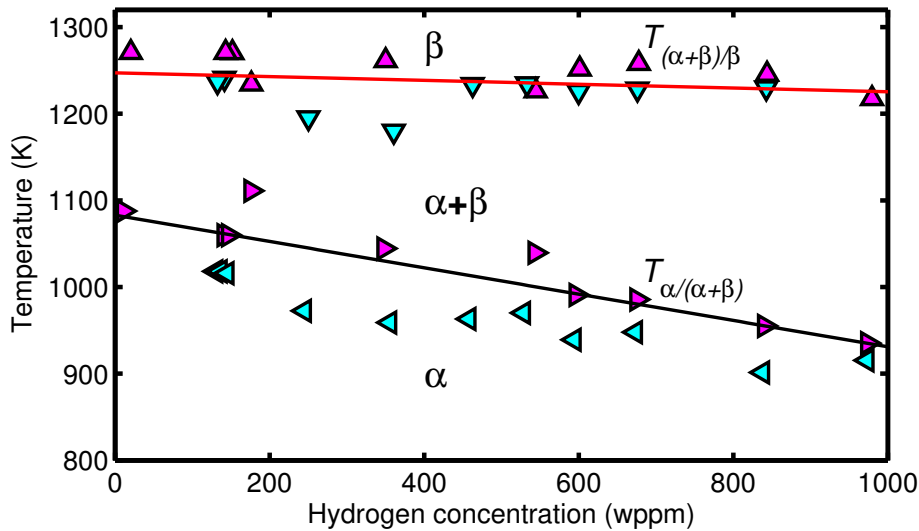}
    \caption{Phase boundary temperatures versus hydrogen concentration for Zircaloy-4L measured upon heating ($\triangle$ {\large$\triangleright$}) and cooling ($\triangledown$ {\large$\triangleleft$}) at  $\approx 0.17$ K/s. The lines (between heating and cooling) are assumed to  represent the equilibrium temperatures; after Brachet et al. \cite{Brachet_et_al_2002}.}
    \label{fig:zry4-h}
  \end{center}
\end{figure}

In a subsequent study, Brachet et al. \cite{Brachet_et_al_2002} investigated the influence of hydrogen content in the alloy on $\alpha/\beta$  phase transformation temperatures of SRA Zircaloy-4 and to some extent  that of RX Zr1NbO. Most of the data presented in \cite{Brachet_et_al_2002} are for Zircaloy-4L (table \ref{tab:chemcomp}) with some discussion on the behavior of Zr1NbO alloy. The data include quasi-equilibrium DSC measurements of the transus temperatures $T_{\alpha/\alpha+\beta}$ and $T_{\alpha+\beta/\beta}$  as a function of hydrogen concentration in the range of [H]$\approx 10$ wppm (as-fabricated value) to  [H]$\approx 970$ wppm; see Fig. \ref{fig:zry4-h}.


No continuous set of data on $\beta$ phase volume fraction versus temperature at different concentrations of hydrogen in Zircaloy-4 is reported in  \cite{Brachet_et_al_2002}. However, computations with some verifications for $\beta$ phase volume fraction versus temperature at hydrogen concentrations of $[H]\approx 0$, 500 and 1000 wppm are reported, which we have used to compare  the results of our model calculations in section \ref{sec:compute}. Regarding kinetic experiments,  the $\alpha/\beta$ phase transformation temperatures were measured by dilatometric technique for Zircaloy-4L during heating at rates of 10 K/s and 100 K/s as a function hydrogen content (in the range 10 - 970 wppm) at $\beta$ volume fractions 0.1, 0.5 and 0.90 \cite{Brachet_et_al_2002}. The measurement uncertainty for the volume fraction and temperature is around $\pm 0.05$ and $\pm 10$ K, respectively.

Frechinet \cite{Frechinet_2001} investigated the $\alpha/\beta$ phase transformation of Zircaloy-4 by electrical resistivity, dilatometry and DSC. His resistivity measurements comprised heating/cooling ratings of $\pm 5$ K/s and $\pm 0.8$ K/s for which the relative amount of $\beta$ phase fraction as a function of temperature was determined. Frechinet's \cite{Frechinet_2001} dilatometric measurements were carried out at a heating rate of 0.8 K/s as a kind of verification of his resistivity measurements. Some discrepancy between the two methods can be observed in regard to the evolution of $\beta$ phase. His DSC measurements covered  heating rates of 0.02, 0.08 and 0.33 K/s, which can be considered as equilibrium or quasi-equilibrium conditions for phase transformation.

A comparative study of the phase transformation behavior of SRA Zircaloy-4 and a RX Zr1NbO alloy at low heating rates is reported in \cite{Kaddour_et_al_2004}. Kaddour et al. \cite{Kaddour_et_al_2004} used both resistivity and DSC to determine the evolution of $\beta$ phase during heating as a function of temperature for Zircaloy-4, in which the heating rates utilized were $\approx 0.8$ K/s (resistivity) and  $\approx 0.3$ K/s (DSC). For Zr1NbO alloy, in addition to resistivity and DSC, they used image analysis, i.e. optical microscopy to determine the volume fraction of $\beta$ phase as a function of temperature. The results obtained by these different techniques are in fair agreement. The onset of the $\alpha \to \beta$ transformation temperature determined by resistivity was about 1093 K for Zircaloy-4 and 1043 K for Zr1NbO alloy.

 More recent investigations of $\alpha \to \beta$ phase transformation kinetics include Nguyen et al's \cite{Nguyen_2017,Nguyen_et_al_2017} in-situ synchrotron X-ray diffraction (SXRD) measurements  of cold-worked (CW) Zircaloy-4L  subjected to heating rates in the range of 10 to 100 K/s, and electrical resistivity measurements of CW Zircaloy-4L and RX Zircaloy-4L at 100 K/s \cite{Nguyen_2017,Nguyen_et_al_2019}, supporting their SXRD measurements. In a new study, Jailin et al. \cite{Jailin_et_al_2020} using dilatometric technique measured the $\alpha \to \beta$ phase transformation kinetics of SRA Zircaloy-4L subjected to very high heating rates, i.e. from 50 to 2000 K/s. They found that from thermal equilibrium to 500 K/s, the temperature at the start of transformation shifts toward higher values with increasing heating rates. However, above 500 K/s, the effect of heating rate saturates. Moreover, they found that the temperature at which the transformation terminates, i.e. when the alloys is in complete $\beta$ phase, is not much affected by  heating rate.

Some  phase transformation kinetic measurements on related Zr-Nb alloys  have also been reported in the literature \cite{Nguyen_et_al_2017,Benes_et_al_2011}.  Nguyen \cite{Nguyen_et_al_2017} has reported in-situ SXRD measurements of  $\alpha$ to $\beta$  phase transformation of RX ZIRLO (table \ref{tab:chemcomp}) at a heating rate of 10 K/s, which indicates that the starting $\alpha \to \beta$ phase transformation temperature  is around 1040 K compared to $\approx$ 1130 K  for CW Zircaloy-4, measured at the same heating rate. However, at about 1170 K the two alloys' $\beta$ fractions become roughly equal, i.e., $\approx 0.4$. B\v{e}nes et al. \cite{Benes_et_al_2011} examined the $\alpha/\beta$ transformation of a type of Zr1NbO alloy, namely the E110 alloy, with an oxygen concentration of about 0.05 wt\% (cf. table \ref{tab:chemcomp}). They measured the temperature of the transformation at low heating rates  in the range of 0.05 to 0.33 K/s by a DSC technique. Their measurements indicate a weak shift (increase)  in the onset of $\alpha$ to ($\alpha+\beta$) transition temperature with heating rate. The equilibrium onset transition temperature for this alloy was put as  $T_{\alpha/(\alpha+\beta)} = 1020$ K.

\section{Models}
\label{sec:models}
In this section, we describe the models that are used to evaluate the experimental data discussed in the preceding section. The models comprise both thermal statics (equilibrium) and kinetics (time varying) for $\alpha \rightleftharpoons \beta$ phase transformation in Zr-alloys (table \ref{tab:chemcomp}) for the intend of computing the  $\beta$ (or $\alpha$) phase volume fractions as a function of time and temperature. Portions of the models were presented in our earlier publications \cite{Massih_Jernkvist_2009,Massih_2009,PM18-001v1}.

\subsection{Thermal equilibrium}
\label{sec:equilib}

We assume that the volume fraction $\beta$-phase material under thermal equilibrium, denoted as $y_\mathrm{eq}$, depends on the Zr-alloy under consideration,
temperature $T$, and excess oxygen and hydrogen concentrations $\bm{x}=(x_\mathrm{O},x_\mathrm{H})$ in the metal \cite{Massih_Jernkvist_2009,Massih_2009}, namely
\begin{equation}
\label{eq:PC_equ}
y_\mathrm{eq}(T,\bm{x}) = \frac{1}{2} \left[ 1 + \tanh{\left( \frac{T-T_\mathrm{mid}(\bm{x})}{T_\mathrm{span}(\bm{x})} \right)} \right],
\end{equation}
where $T_\mathrm{mid}$ and $T_\mathrm{span}$ are material specific functions of the cladding metal layer excess oxygen concentration (>1200 wppm), and hydrogen concentration (> 10 wppm). They are related to the middle and the span of the mixed-phase temperature region at equilibrium. They are calculated from the mixed phase lower and upper boundary temperatures, $T_{\alpha}$ and $T_{\beta}$, through
\begin{equation}
\label{eq:T_c}
T_\mathrm{mid}(\bm{x}) = \frac{T_{\alpha}(\bm{x}) + T_{\beta}(\bm{x})}{2},
\end{equation}
\begin{equation}
\label{eq:T_s}
T_\mathrm{span}(\bm{x}) = \frac{T_{\beta}(\bm{x}) - T_\mathrm{mid}(\bm{x})}{2.3}.
\end{equation}
For the cladding materials considered (table \ref{tab:chemcomp}), the phase boundary temperatures $T_{\alpha}$ and $T_{\beta}$ can
be written as
\begin{equation}
\label{eq:T_phases}
T_{\alpha / \beta}(\bm{x}) = T_{0} + A_1 \: x_\mathrm{O}^{m} - A_2 x_\mathrm{H},
\end{equation}
where $x_\mathrm{O}$  and  $x_\mathrm{H}$ are the weight fraction of excess oxygen and hydrogen in the cladding metal respectively,  and $A_1$, $A_2$ and $m$ are constants given in table \ref{tab:Phase_boundaries}. Equation \ref{eq:T_phases} treats  $x_\mathrm{O}$  and  $x_\mathrm{H}$ as independent variables with no interaction between them. Furthermore, the relation is appropriate for $x_\mathrm{O}<0.01$  and  $x_\mathrm{H}<0.001$. The phase boundary temperatures versus excess oxygen concentration, calculated through Eq. (\ref{eq:T_phases}), are compared with experimental data for Zircaloy-4, Zr-0.5\%NbO and Zr-1\%Nb in Fig. \ref{fig:zry-o}.

\begin{table}[ht]
\centering
\begin{tabular}{|l|c|c|c|c|c|}
\hline
  Alloy & Phase boundary &  $T_{0}$  &        $A_1$    & $A_2$  &  $m$  \\
   type & temperature    &   (K)  &         (K)             & (K)      & (-)       \\
\hline
  \multirow{2}{*}{Zircaloy-4} & $T_{\alpha}$     &   1075   & 2.713$\times$10$^{5}$ & $1.52\times$10$^{5}$ &   1.732 \\
                              & $T_{\beta}$      &   1250   & 3.138$\times$10$^{4}$ & $2.20\times$10$^{4}$ &   0.929 \\
 \hline
  \multirow{2}{*}{Zr1NbO}   & $T_{\alpha}$     &   1016   & 2.713$\times$10$^{5}$ & ----- & 1.732 \\
                              & $T_{\beta}$      &   1240   & 3.138$\times$10$^{4}$ & ----- & 0.929 \\
\hline
  \end{tabular}
  \caption{\label{tab:Phase_boundaries}Constants for calculating phase boundary temperatures of Zircaloy-4 and Zr1NbO cladding through Eq. (\ref{eq:T_phases}).}
\end{table}

\subsection{Kinetics}
\label{sec:kinetics}

Our approach to kinetics of the $\alpha/\beta$ phase transformation in zirconium alloys is similar to the method used by Leblond and Devaux \cite{Leblond_Devaux_1984} for diffusion controlled transformations in steels. We assume that there are only two phases, $\alpha$ and $\beta$, and we denote the volume fraction of phase $\beta$ by the variable $y$. We also posit  unique phase transformations $\alpha\rightleftharpoons(\alpha+\beta)\rightleftharpoons\beta$ occurring at specified temperatures. Furthermore, we assume that the kinetics equation furnishes the equilibrium proportion $y_\mathrm{eq}$ discussed in the foregoing section.

The general form of the kinetics or the time derivative of $y$, $\dot{y}=dy/dt$, depends only on $y$ and $T$, namely
\begin{equation}
\label{eq:gen-kin}
\dot{y} = f(T,y),
\end{equation}
where $f$ is a function of thermal history $T(t)$ and $y$ is the volume fraction of $\beta$ phase, and in thermal equilibrium $f(T,y=y_\mathrm{eq})=0$.
In modeling  the kinetics of the $\alpha/\beta$ transformation, two variants of the model are presented here, which we call model A and model B. In model A, a key parameter entering the kinetic equation is temperature rate of change ($q=dT/dt$), i.e. in principle $\dot{y}=f(q,T,y)$, whereas in model B, $q$ does not appear explicitly in this equation. In the forthcoming two subsections, we will delineate the equations used for these two models.

\subsubsection*{Model A}
\label{sec:modA}

We consider a simplest equation of type \eqref{eq:gen-kin}, by assuming $y=y_\mathrm{eq}(T)$ is a steady-state solution at constant temperature,  i.e. $f(T,y=y_\mathrm{eq})=0$ for each $T$. We obtain $f$ by assuming $y$ is not too far from thermal equilibrium; thereby we Taylor expand $f$ up to the first power
\begin{equation}
\label{eq:taylor-expand}
f(T,y) \approx f\big(T,y_\mathrm{eq}(T)\big) + \frac{\partial f}{\partial y}\Big\arrowvert_{y=y_\mathrm{eq}}\big[y-y_\mathrm{eq}(T)\big].
\end{equation}
Since the first term on the right-hand side is null, we can write Eq. (\ref{eq:gen-kin}) as
\begin{equation}
\label{eq:moda-kin}
\dot{y}  = \frac{1}{\tau(T)}\Big[y_\mathrm{eq}(T)-y\Big],
\end{equation}
where $\tau$ is a relaxation time, which is temperature dependent and formally is defined as
\begin{equation}
\label{eq:tau}
\frac{1}{\tau(T)}  = -\frac{\partial f}{\partial y}\Big\arrowvert_{y=y_\mathrm{eq}}.
\end{equation}
$\tau(T)$ is a positive quantity and  characterizes the duration of phase transformation. Equation \eqref{eq:moda-kin} assumes initial values $y(0)=0$ for heating and $y(0)=1$ for cooling. For  model A during heat-up, $\tau$ is calculated through an empirical relation by \cite{Massih_2009}
\begin{equation}
\label{eq:chartime}
\tau = \frac{\exp\big[16650/T\big]}{60457+18129q},
\end{equation}
where $\tau$ is in s, $T$ is in kelvin, and $q=dT/dt$ is the heating rate in K/s. Furthermore,  $y_\mathrm{eq}(T)$ is given by Eq. \eqref{eq:PC_equ}.

During cooling  ($\beta \to \alpha$), however, as the temperature is lowered ($q<0$), $\tau$ rapidly increases using this formula. And at sufficiently low temperatures, it blows up thus rendering artificially $\dot{y}=0$ in Eq. \eqref{eq:moda-kin}. Surely the formula \eqref{eq:chartime} requires a constraint during cooling to get meaningful results. Hence, in our computer program, we have modified Eq. \eqref{eq:chartime} according to
\begin{equation}
\label{eq:chartime-cool}
\tau_{cool} = \frac{\exp\big[16650/(T+T_\mathrm{b})\big]}{60457+18129|q|},
\end{equation}
that is, we have introduced a lower-bound temperature constraint $T_\mathrm{b}$ to avoid unphysically slow phase transformation during rapid cooling to low temperature. Choosing $T_\mathrm{b}=1000$ K, gives reasonable results for experiments performed with cooling rates from $-5$ to $-100$ K/s;  other choices are also possible. In model A, the heating/cooling rate $q$ also enters the expressions used for the phase boundary temperatures $T_{\alpha}$ and $T_{\beta}$ \cite{Massih_2009}.

Model A may be used for an experiment or a fabrication process when the heating/cooling rate $q$ is prescribed in advance. As will be shown in section \ref{sec:comp-moda}, the model describes  measured data satisfactorily, i.e. it  is engineered to do so.
\subsubsection*{Model B}
\label{sec:modB}

As noted in the foregoing subsection, the relaxation time in model A embraces the temperature rate $q$, thereby renders it ineffective for reactor fuel rod simulations, where temperature history is not known a priori and  may vary nonuniformly and rapidly during transients. In order to remedy this shortcoming, in model B we have extended equation \eqref{eq:moda-kin} to a higher order  and modified the relaxation time $\tau$ in Model A to fit the measured data suitably. Moreover, we have introduced a cutoff on $\tau$.

The basic equations for the volume fraction of the $\beta$ phase in model B are
\begin{equation}
\label{eq:modb-kin-hc}
\frac{d y}{d t} = \frac{1}{\tau(T)}\Big((y_\mathrm{eq}(T)-y) \pm \frac{b}{y_\mathrm{eq}(T)}\big(y_\mathrm{eq}(T)-y\big)^2\Big),
\end{equation}
where the +sign is for heat-up ($y<y_\mathrm{eq}$) while  the $-$sign is for cool-down ($y>y_\mathrm{eq}$), and $y_\mathrm{eq}(T)$ is given by Eq. \eqref{eq:PC_equ}. The relaxation time $\tau$  (in s) is given as
\begin{equation}
\label{eq:chartime-2}
\tau = B\exp(E/T),
\end{equation}
where $B$,  $b$ and $E$ are constants; see table  \ref{tab:relax-time}. During cooling, we have introduced an upper cutoff limit on time constant $\tau\leq \tau^\ast$ to avoid excessively long relaxation times, namely we take
\begin{equation}
\label{eq:cutoff}
\tau = \min(\tau^\ast,B \, e^{E/T}),
\end{equation}
where $\tau^\ast=6$ s is used in all our computations (table \ref{tab:relax-time}). It will be shown in section \ref{sec:comp-modb} that model B reproduces measured data fairly satisfactorily.

\begin{table}
\centering
 \begin{tabular}{|l|c|c|c|c|}
  \hline
  Alloy type &   $E$  &        $B$    & $b$  &  $\tau^\ast$  \\
            &   (K)     &         (s)             & (-)      & (s)       \\
 \hline
  Zircaloy-4     &  54000   & $1.05 \times 10^{-19}$ & 0.3 &   6.0 \\
 \hline
 Zr1NbO         &   54000   & $1.55 \times 10^{-19}$ & 0.3 & 6.0 \\
\hline
  \end{tabular}
\caption{\label{tab:relax-time}Model B parameter constants, Eqs. \eqref{eq:modb-kin-hc}-\eqref{eq:cutoff}.}
\end{table}

\section{Results: Computations vs. measurements}
\label{sec:compute}
In this section, we use the models described in the foregoing section to evaluate some of the data alluded in section \ref{sec:expdata}. We compare model computations of  the volume fraction of $\beta$ phase with experimental data  through a number of figures.

\subsection{Thermal equilibrium}
\label{sec:equilib-comp}

\begin{figure}[!hbtp]
  \begin{center}
    \includegraphics[width=0.7\textwidth]{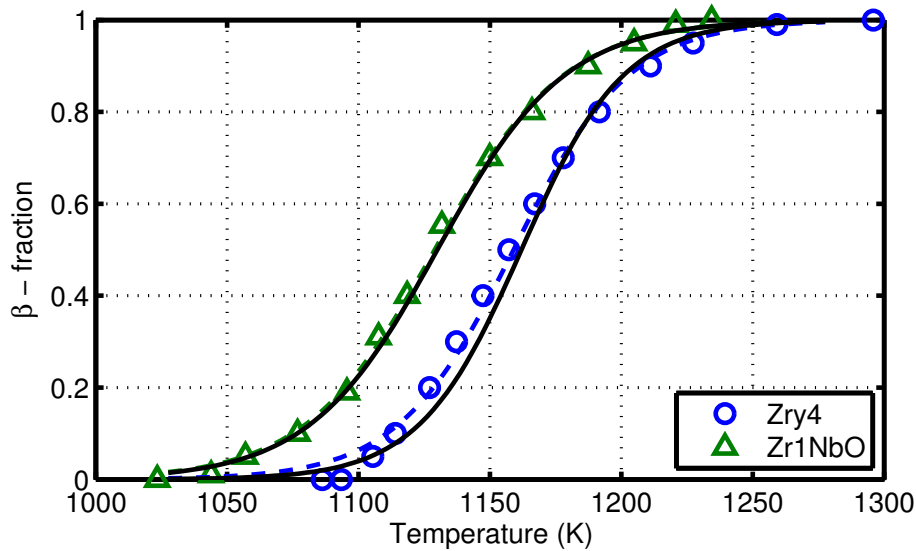}
    \caption{Equilibrium volume fraction of $\beta$ phase as a function of temperature calculated (lines) vs. measured data (markers)  \cite{Forgeron_et_al_2000} for Zircaloy-4 and Zr-1NbO alloy. Solid lines are calculated with the model parameters listed in table  \ref{tab:Phase_boundaries} whereas the dashed lines are those from ref. \cite{Massih_2009}.}
    \label{fig:zreq}
  \end{center}
\end{figure}

In Fig. \ref{fig:zreq},  the temperature dependence of the equilibrium volume fraction of the $\beta$ phase Zircaloy-4 and Zr1NbO alloy using Eq. \eqref{eq:PC_equ} is shown against the experimental data (markers) extracted from ref. \cite{Forgeron_et_al_2000}. Solid lines are calculated with the model parameters listed in table  \ref{tab:Phase_boundaries} whereas the dashed lines are those from ref. \cite{Massih_2009}; here $x_{\mathrm{O}}=x_{\mathrm{H}}=0$. As can be seen, the agreement is quite good. It is noted that the transition temperatures for Zr1NbO are lower than for Zircaloy-4 and  the volume fraction of $\beta$ phase shifts to a lower temperature because niobium acts as a $\beta$ phase stabilizer.

\begin{figure}[!hbtp]
  \begin{center}
    \includegraphics[width=0.7\textwidth]{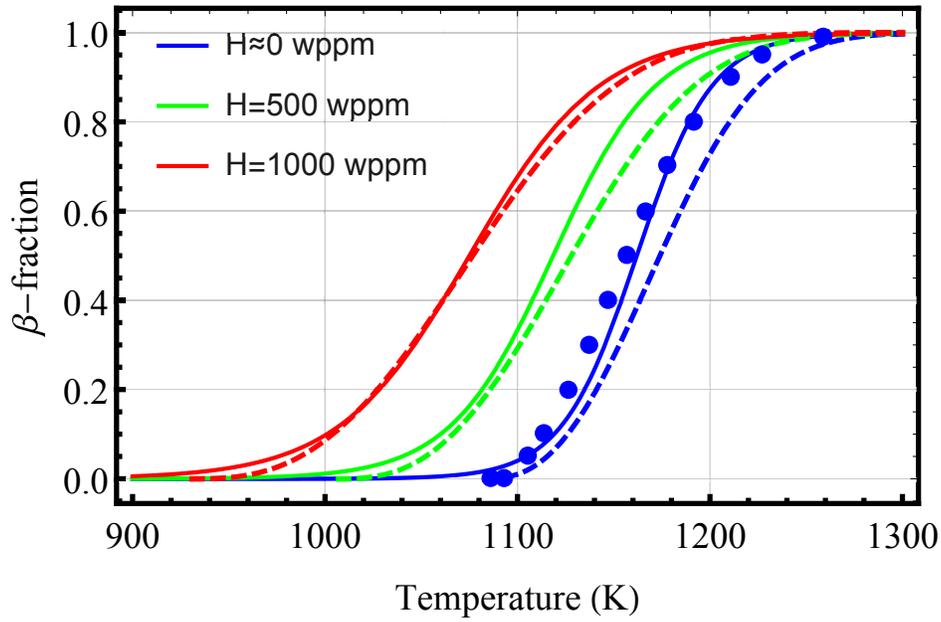}
    \caption{Equilibrium volume fraction of $\beta$ phase as a function of temperature calculated  for Zircaloy-4 with several concentrations of hydrogen. Solid lines are calculated with model parameters listed in table  \ref{tab:Phase_boundaries} whereas the dashed lines are those extracted from figure 12 of ref. \cite{Brachet_et_al_2002}, which are based on analysis of experimental data. The disks show measured data for $\text{H}\approx 0$ wppm from ref. \cite{Forgeron_et_al_2000}.}
    \label{fig:zry4-eqh}
  \end{center}
\end{figure}

The influence of hydrogen content on phase transformation of Zircaloy-4 is displayed in Fig.  \ref{fig:zry4-eqh}. It is seen that the results of our computations for hydrogenated Zircaloy-4 are in fair agreement with experimental data and analysis in \cite{Brachet_et_al_2002}. However, the deviation for as-fabricated Zircaloy-4 (H$\approx 0$ wppm) is salient.  This is because, our equilibrium model is calibrated with the data in \cite{Forgeron_et_al_2000}; discs in the figure. Note that as hydrogen concentration is increased, the $\alpha \to (\alpha+\beta)$ transition temperature shifts to a lower value and the volume fraction of the favored phase is moved to lower temperatures. We should note that, to our knowledge, no similar measurements on the effect of excess oxygen on phase transformation of zirconium alloys have been reported in the literature.


\subsection{Kinetics}
\label{sec:comp-kinetics}

\subsubsection*{Model A}
\label{sec:comp-moda}

The results of model A computations were presented earlier  \cite{Massih_2009}, however, with some clarification included here that was missing in that reference; e.g. usage of  Eq. \eqref{eq:chartime-cool}, and additional data. We perform our computations using this model and parameter values as in  \cite{Massih_2009} for the sake of a benchmark with model B, which is the main contribution of our paper.

\begin{figure}[htbp]
\begin{center}
    \begin{tabular}{ll}
    \includegraphics[width=0.7\textwidth]{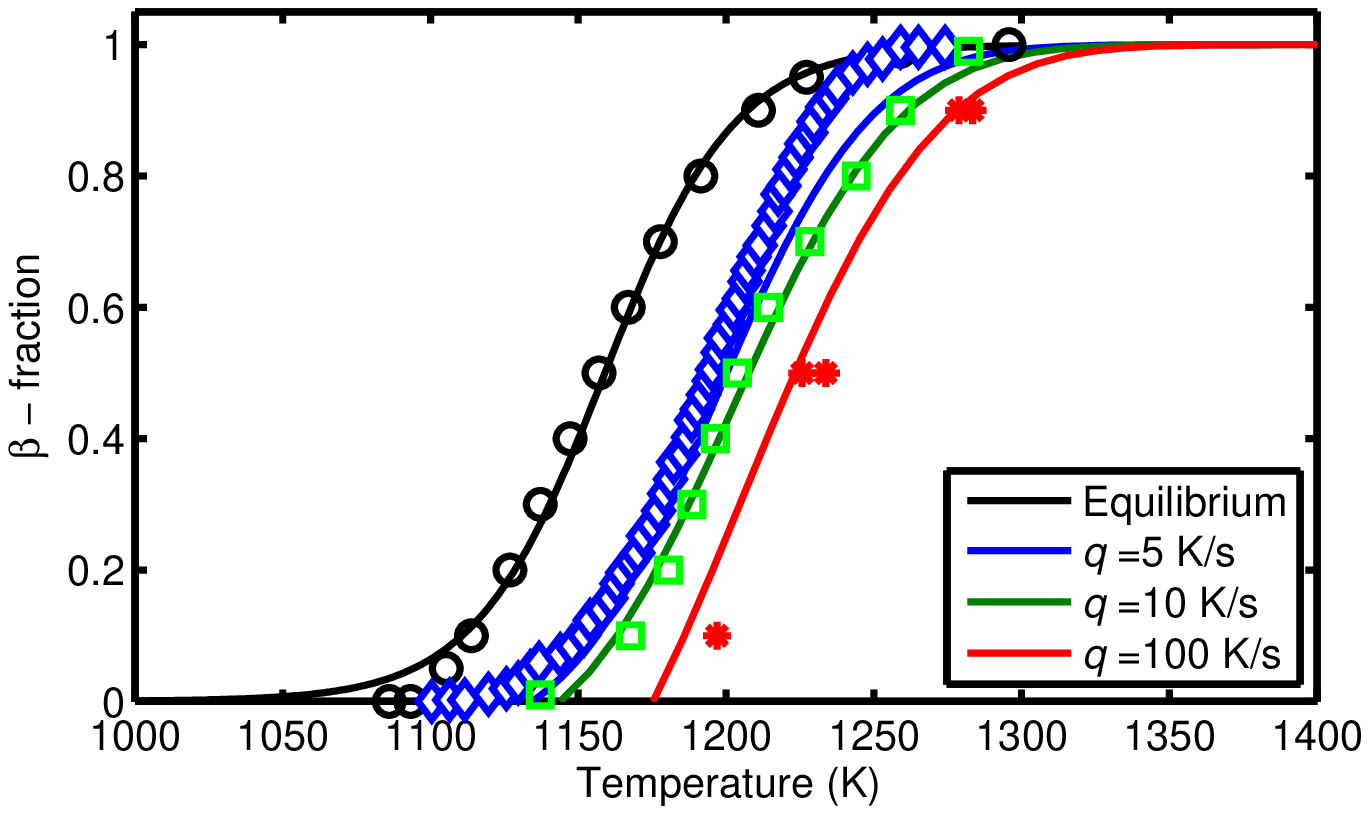}\\
   \vspace{12pt}
    \includegraphics[width=0.7\textwidth]{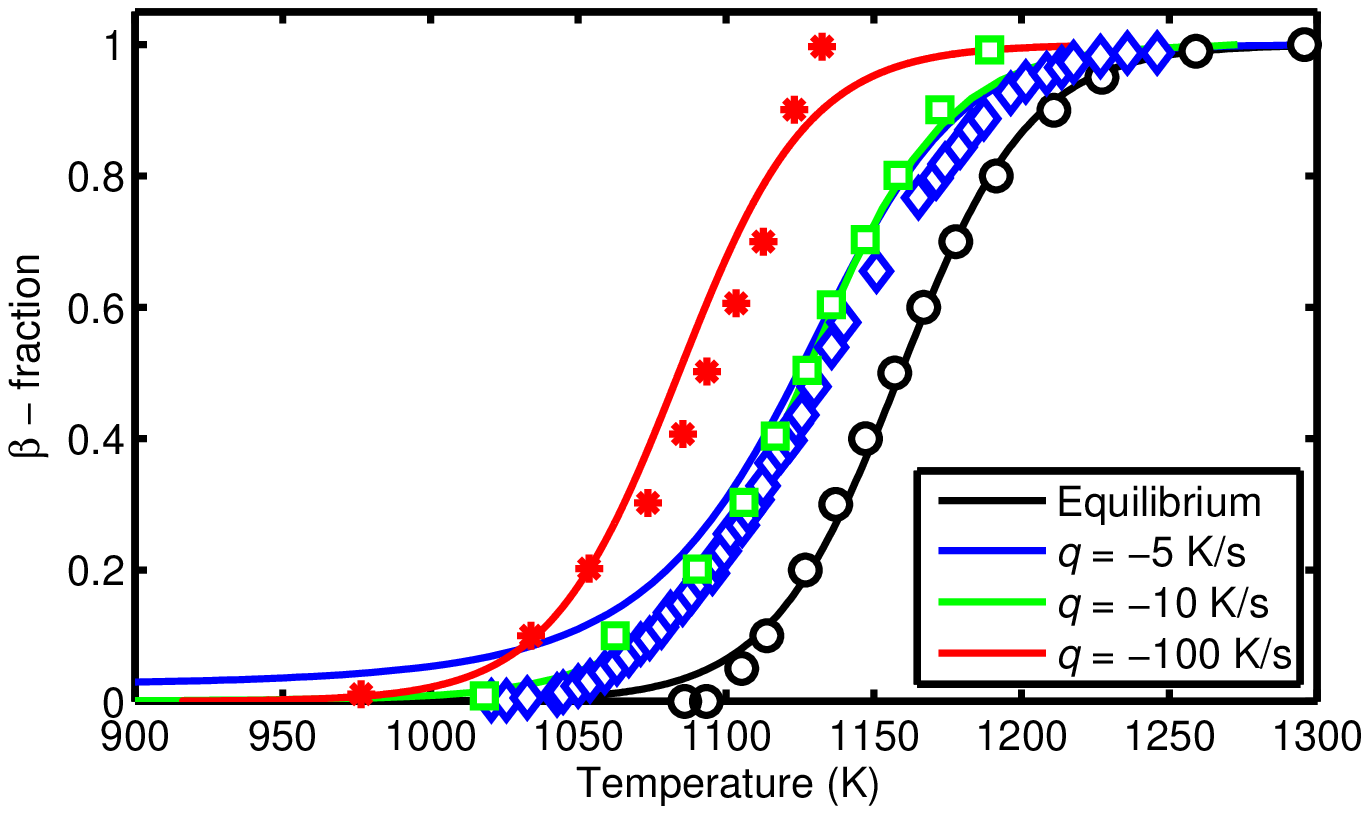}\\
\end{tabular}
\caption{Volume fraction of $\beta$ phase as a function of temperature calculated (lines) by using model A vs. measured data (markers)  \cite{Forgeron_et_al_2000,Frechinet_2001} for Zircaloy-4 at different heating/cooling rates (top/bottom) where  $q=dT/dt$.}
\label{fig:beta-jnm08}
  \end{center}
\end{figure}

\begin{figure}[!hbtp]
  \begin{center}
    \includegraphics[width=0.70\textwidth]{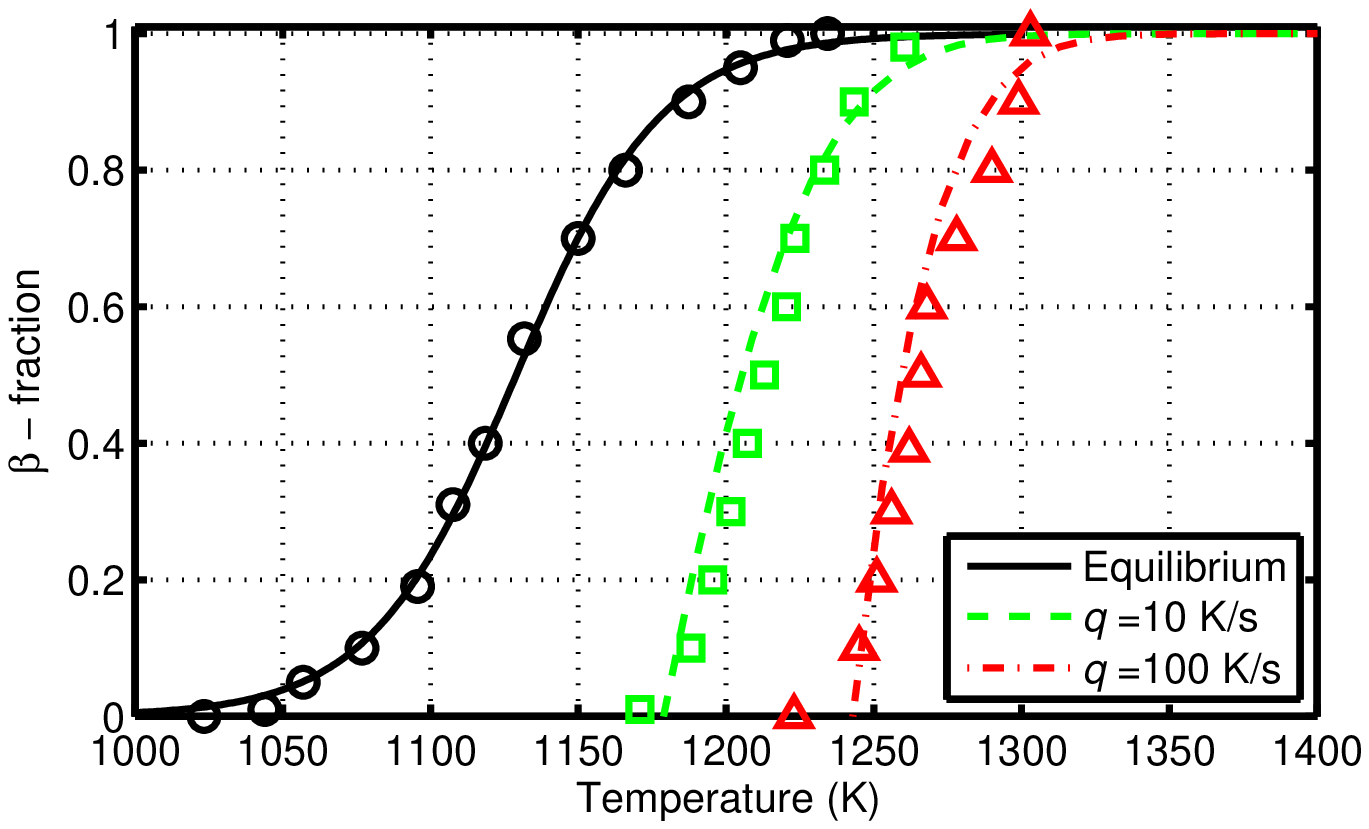}
    \caption{Volume fraction of $\beta$ phase as a function of temperature calculated (lines) by using model A vs. measured data (markers)  \cite{Forgeron_et_al_2000} for Zr1NbO at heating rates $q=dT/dt=10$ and 100 K/s and at thermal equilibrium.}
    \label{fig:m5-heat2}
  \end{center}
\end{figure}

 We evaluate some of the measurements discussed in section \ref{sec:expdata}, namely data  in \cite{Forgeron_et_al_2000} with heating/cooling rates $q=dT/dt=\pm 10, \pm 100$ K/s and in \cite{Frechinet_2001}  with  $q=dT/dt=\pm 5$ K/s. We solve Eq. \eqref{eq:moda-kin} with initial conditions $y(0)=0$ on heating and $y(0) = 1$ on cooling together with relations \eqref{eq:chartime}-\eqref{eq:chartime-cool}  for the relaxation times (with $T_\mathrm b=25|q|$) by using an explicit Runge-Kutta method. The results for the fraction of the transformed phase for Zircaloy-4 under heating/cooling are shown in Fig. \ref{fig:beta-jnm08}. The markers denote measured data while the lines are calculations according to model A. The corresponding calculations and data for Zr1NbO \cite{Forgeron_et_al_2000} subject to heating rates $q=10$ and $100$ K/s are shown in Fig. \ref{fig:m5-heat2}. It is seen that as $q$ is increased, the transformed volume fraction shifts to higher/lower temperatures under heating/cooling, respectively.
 

\subsubsection*{Model B}
\label{sec:comp-modb}

\begin{figure}[t]
\centering
\subfigure[]{\label{fig:beta-heat-loj2}\includegraphics[scale=0.80]{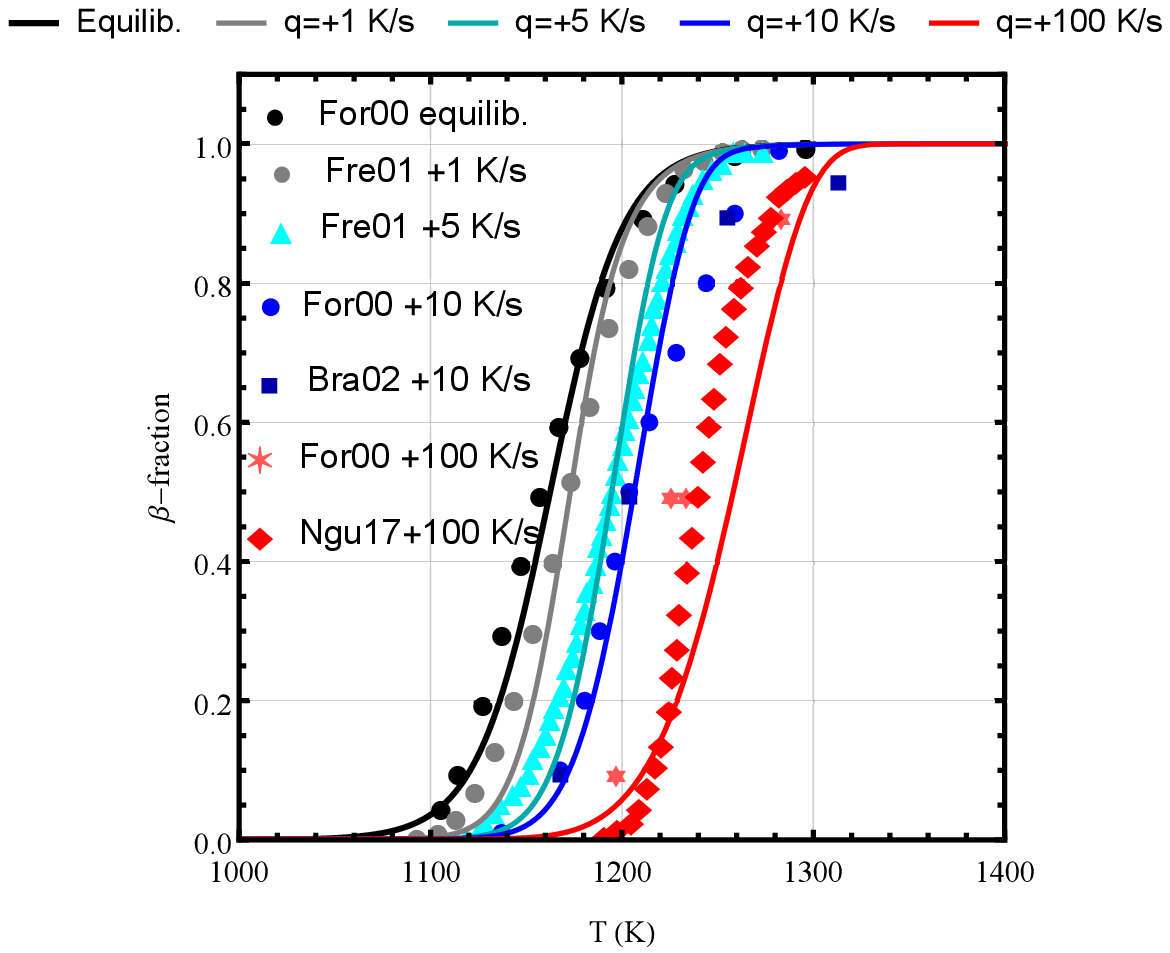}}
\subfigure[]{\label{fig:beta-cool-loj2}\includegraphics[scale=0.80]{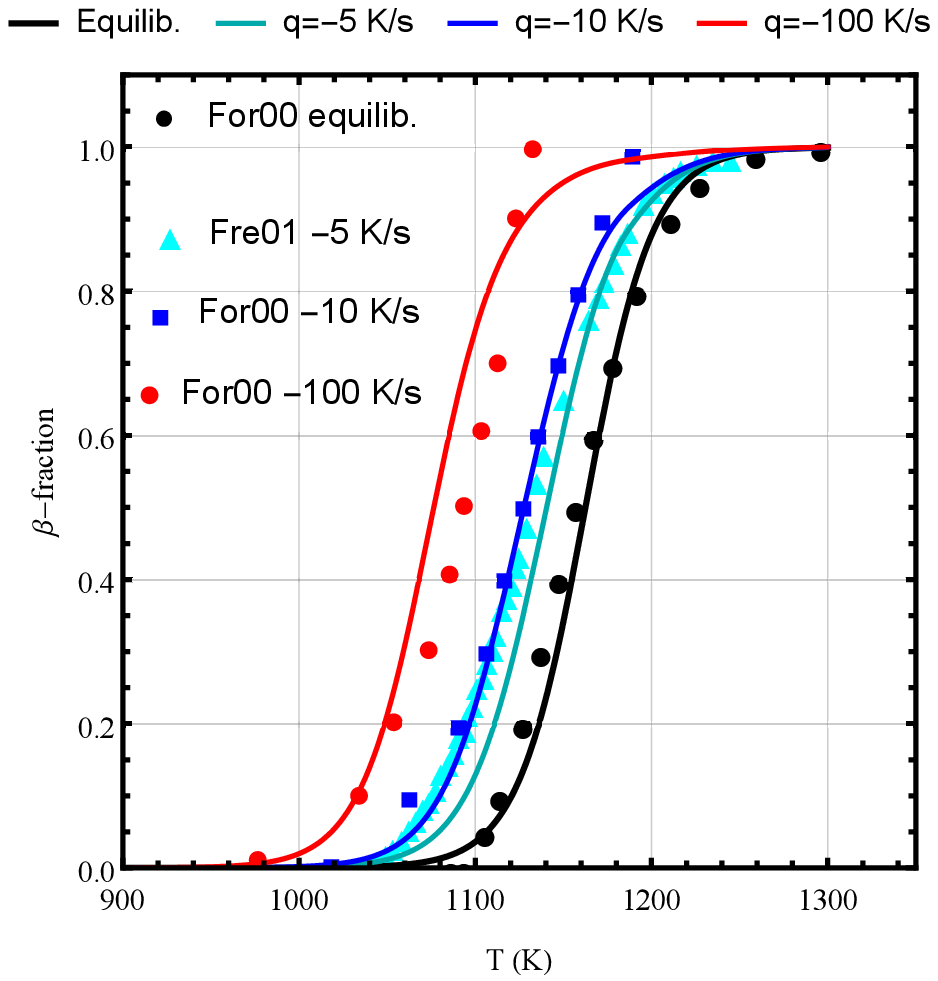}}
\caption{Volume fraction of $\beta$ phase as a function of temperature calculated (lines) by using model B vs. measured data (markers) for Zircaloy-4 at different heating/cooling rates (left/right panel) where $q=dT/dt$. Measured data refs. are For00\cite{Forgeron_et_al_2000}, Fre01\cite{Frechinet_2001}, Bra02\cite{Brachet_et_al_2002}, and Ngu17\cite{Nguyen_2017}.}
\label{fig:beta-loj2}
\end{figure}

\begin{figure}[!hbtp]
  \begin{center}
    \includegraphics[width=0.60\textwidth]{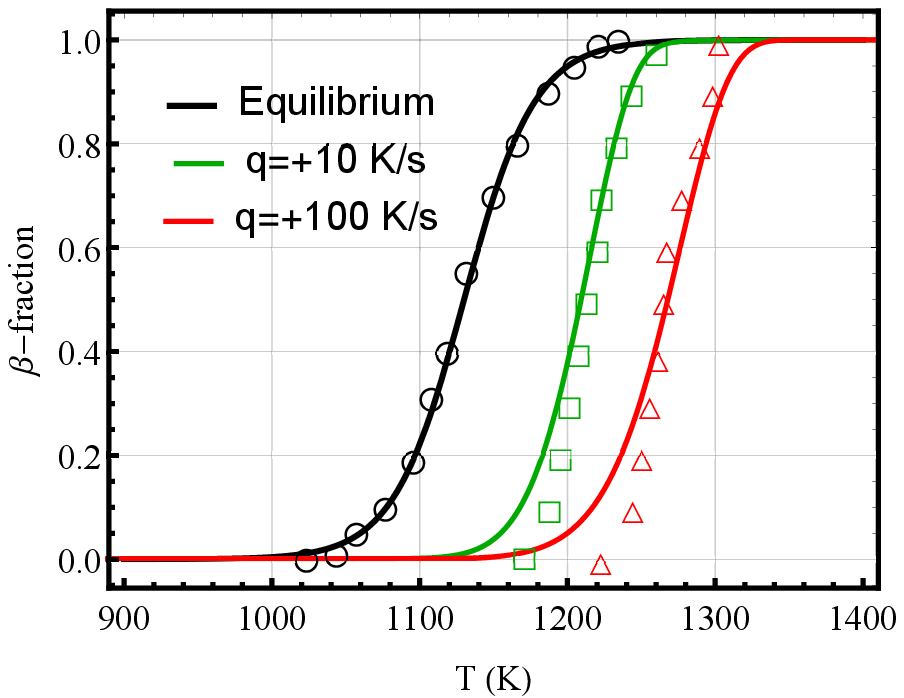}
    \caption{Volume fraction of $\beta$ phase as a function of temperature calculated (lines) using model B vs. measured data (markers)  \cite{Forgeron_et_al_2000} for Zr1NbO at heating rates $q=dT/dt=10$ and 100 K/s and at thermal equilibrium.}
    \label{fig:beta-heat-m5}
  \end{center}
\end{figure}

Here, we solve Eq. \eqref{eq:modb-kin-hc}  with initial conditions $y(0)=0$ on heating and $y(0) = 1$ on cooling together with relations \eqref{eq:chartime-2}-\eqref{eq:cutoff}  for the relaxation time by using an explicit Runge-Kutta method. The results of model B computations, the volume fraction of $\beta$ phase versus temperature, against measured data for Zircaloy-4 (markers) which were performed at thermal rates of $\pm 0.8 (\approx \pm 1) \pm 5,\pm10,\pm 100$ K/s and at thermal equilibrium (cf. section \ref{sec:expdata}) are shown in Fig.  \ref{fig:beta-loj2}. As can be seen from this figure, the agreement between calculations and measurements are satisfactory, except for the case of heating at 100 K/s, at which computations underestimate the measurements at temperatures $\approx 1230$ and $\approx 1280$ K. However, the trend in  $\alpha \to \beta$  phase transition is qualitatively featured. The corresponding calculations and data for Zr1NbO \cite{Forgeron_et_al_2000} subject to heating rates of $+10$, $+100$ K/s and at thermal equilibrium are plotted in Fig.  \ref{fig:beta-heat-m5};  cf. Fig. \ref{fig:m5-heat2}. As can be seen, the concordance between computations and measurements is fair for this alloy.

\begin{figure}[!hbtp]
  \begin{center}
    \includegraphics[width=0.90\textwidth]{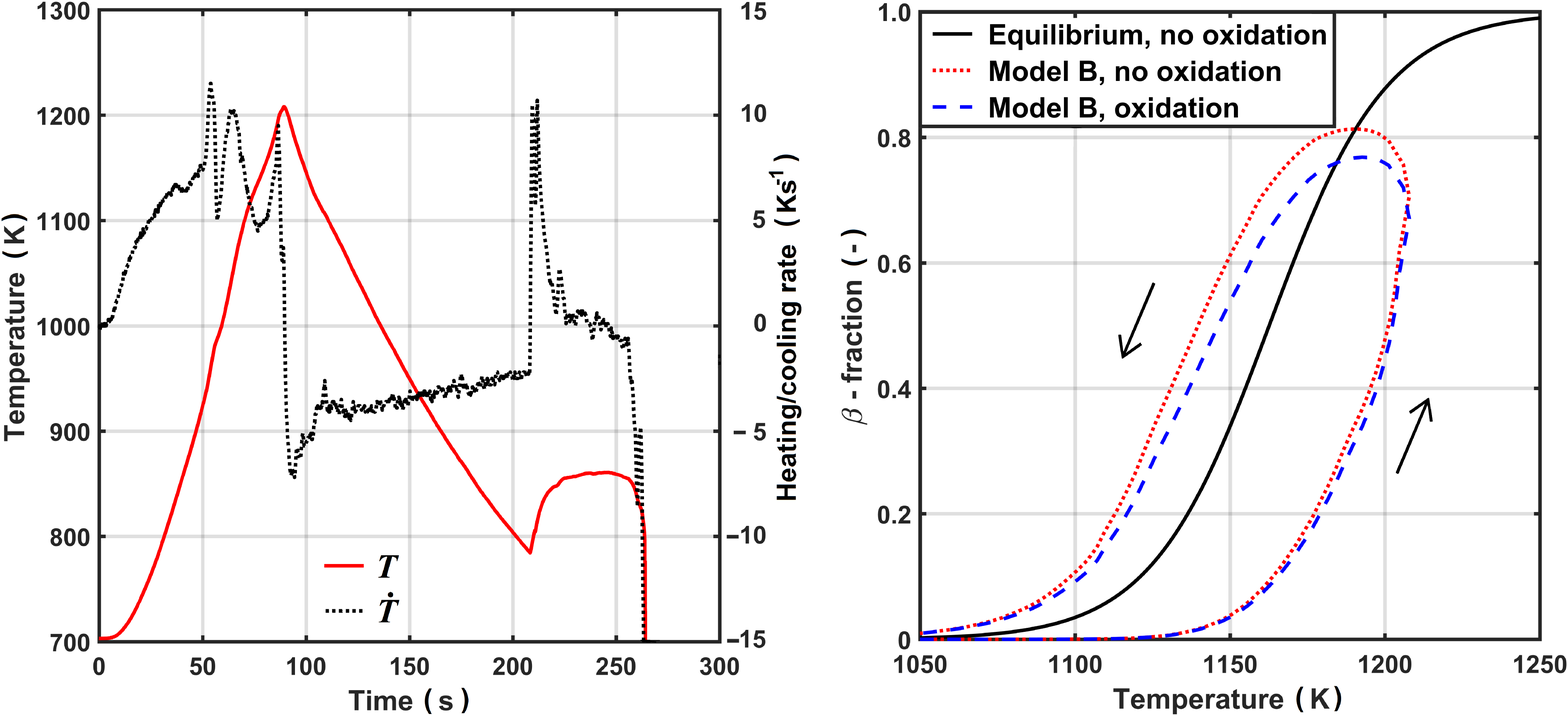}
    \caption{Left panel: Measured temperature history from the QUENCH-L1 LOCA test considered in calculations. Right panel: Calculated $\beta$-fraction in Zircaloy-4 cladding for the simulated temperature history, with and without consideration of cladding high-temperature oxidation. The equilibrium curve from Fig. \ref{fig:zreq} is included for comparison.}
    \label{fig:QL1}
  \end{center}
\end{figure}

To illustrate the phase transformation kinetics expected for reactor fuel cladding under a postulated loss-of-coolant accident (LOCA) in a pressurized water reactor (PWR) \cite{NEA_LOCA_SOAR_2009}, we apply model B to a cladding temperature history measured in a test within the QUENCH-LOCA experimental program at the Karlsruhe Institute of Technology, Germany. This program comprised seven out-of-reactor LOCA-simulation tests on bundles with 21 electrically heated and extensively instrumented fuel rod simulators with various Zr-base cladding materials \cite{Stuckert_et_al_2020}. The rods in the considered test, QUENCH-L1, were clad with as-received Zircaloy-4 \cite{Stuckert_et_al_2018}. More specifically, we consider the cladding temperature history measured for rod 4 at an elevation 450 mm from the bottom of the heated section (thermocouple TFS 4/8 \cite{Stuckert_et_al_2018}), which is shown in the left panel of Fig. \ref{fig:QL1}. The temperature history is typical for a postulated large-break LOCA in a PWR: An initial heat-up phase, representing loss of primary coolant water at full reactor power, is followed by cooling in steam with a rate around 3-4 Ks$^{-1}$. At $t$=207 s, the axial steam flow through the fuel bundle is blocked by liquid water that reaches the lower (cold) part of the bundle during re-flood of the reactor pressure vessel by the emergency core cooling system. The fuel cladding temperature then rises to about 870 K until the water front reaches the thermocouple position and the cladding is rapidly quenched.

For illustration, this temperature history is simulated with model B. The calculated results are shown in the right panel of Fig. \ref{fig:QL1} for two different cases: For the 'no oxidation' case, the effects of high-temperature oxidation are neglected, keeping $x_{\mathrm{O}}$=$x_{\mathrm{H}}$=0 throughout the simulated test. For the 'oxidation' case, the evolution of $x_\mathrm{O}$ during the test is calculated through the correlation by Leistikow and Schanz \cite{Leistikow_Schanz_1987}, while hydrogen pick-up during the test is neglected ($x_{\mathrm{H}}$=0). The calculated results in Fig. \ref{fig:QL1} suggest significant kinetic effects on the cladding phase composition during a typical PWR large-break LOCA. The calculated effects of cladding high-temperature oxidation are, however, moderate. The calculated excess oxygen concentration in the cladding metal is merely 280 wppm at end of the simulated test. High-temperature (600--1200$^\circ$C) oxidation kinetics and hydrogen pickup of zirconium alloys, important during LOCA conditions, have been extensively studied in the literature over the years. Some recent studies include refs. \cite{Stuckert_et_al_2020,chuto2009high,steinbruck2011oxidation}.

\begin{figure}[t]
\centering
\subfigure[]{\label{fig:BetaHeat-htField100m}\includegraphics[scale=0.65]{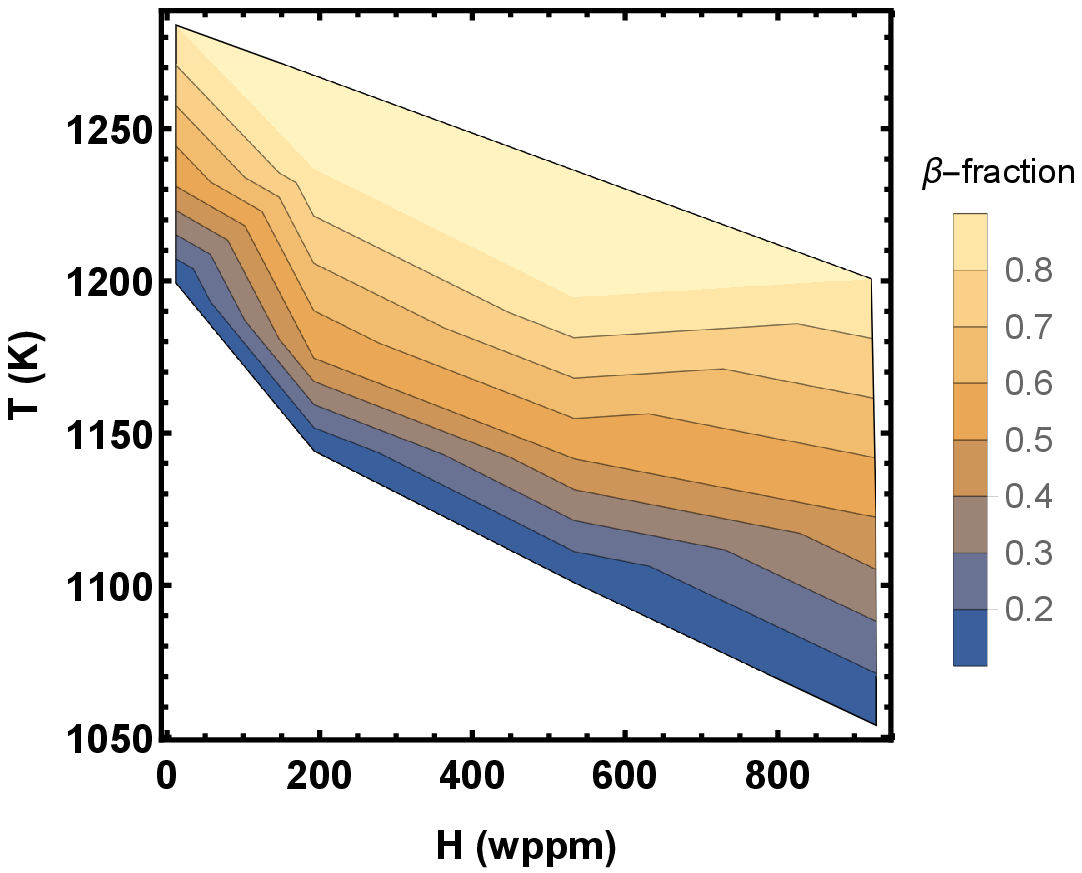}}
\subfigure[]{\label{fig:BetaHeat-htField100c}\includegraphics[scale=0.65]{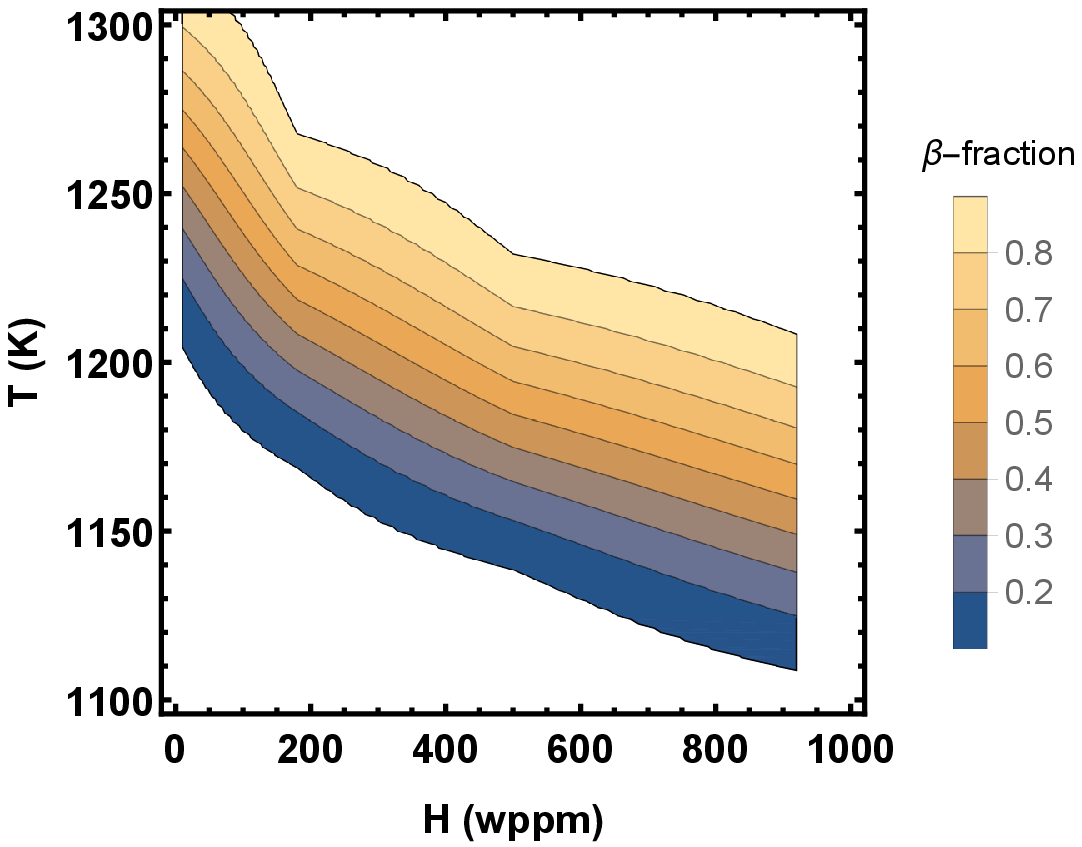}}
\caption{Contour plots of $\alpha/\beta$ phase transformation temperatures in Zircaloy-4 as a function of hydrogen concentration. (a) Measured on heating at 100 K/s,  (b)  calculated (model B) on heating at 100 K/s. Measurements are based on the data reported in \cite{Brachet_et_al_2002}.}
\label{fig:BetaHeat_htField}
\end{figure}

Next, we consider the case of hydrogenated Zircaloy-4 and rely on experimental data reported in \cite{Brachet_et_al_2002}. As mentioned in section \ref{sec:expdata}, no continuous set of data on $\beta$ phase volume fraction versus temperature at different concentrations of hydrogen in Zircaloy-4 have been reported in  \cite{Brachet_et_al_2002}. Brachet et al. \cite{Brachet_et_al_2002}, besides equilibrium data, have presented data on heating at 10 K/s and 100 K/s for $\beta$ phase volume fractions of 0.1, 0.5 and 0.9 for Zircaloy-4 containing hydrogen concentrations in the range $\approx$ 10 to 970 wppm. We have extracted these data from their figure 8 by grouping the data to four levels of hydrogen concentrations, namely,  $\approx$ 12 wppm, 188 wppm, 500 wppm, and 920 wppm. The results are depicted as a contour plot in Fig.  \ref{fig:BetaHeat-htField100m} for the heating rate of 100 K. In this figure, the $\beta$-fraction values  are interpolated  between the measured data points of  0.1, 0.5 and 0.9 for the sake of contiguity. The corresponding computations with model B are presented in Fig.  \ref{fig:BetaHeat-htField100c}. It is seen that measurements and computations are in fair agreement, albeit the complexity of the phenomenon.

In order to recount the aforementioned experimental data on Zircaloy4+H, we have extended the relaxation time relation of phase transformation, i.e. Eq. \eqref{eq:chartime-2}, to account for hydrogen concentration as
\begin{equation}
\label{eq:chartime-2h}
\tau =\frac{ B\exp(E/T)}{1-0.0003 \bar x_\mathrm{H}\big(1-\exp(5000/T)\big)}.
\end{equation}
 Here, $\bar x_\mathrm{H}$ is hydrogen concentration in metal in wppm. This modification is similar to that of extending the diffusion coefficient of hydrogen in a metal in the presence of hydrogen traps \cite{Hagi_Hayashi_1987}. Relation \eqref{eq:chartime-2h} should give plausible results for hydrogen concentrations of up to 1000 wppm in Zircaloy-4 in $T$-range $\approx 1000-1600$ K.

\begin{figure}[t]
\centering
\subfigure[]{\label{fig:BetaHeat-ht10c}\includegraphics[scale=0.70]{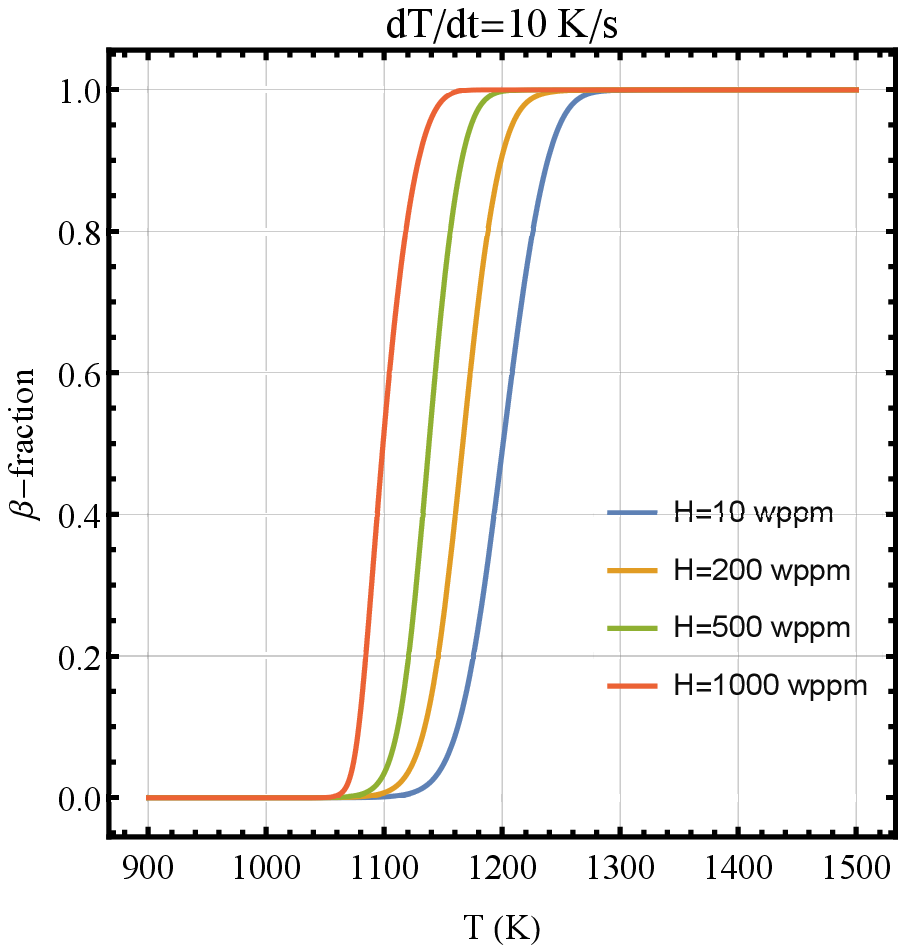}}
\subfigure[]{\label{fig:BetaHeat-100c}\includegraphics[scale=0.70]{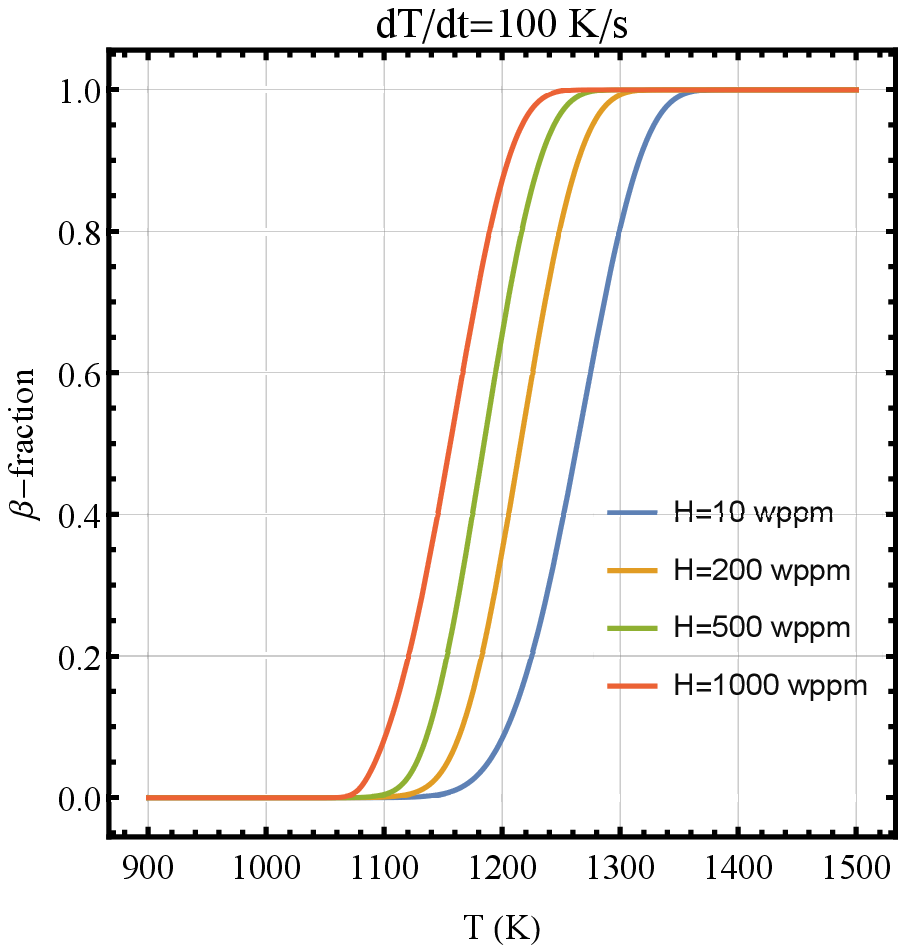}}
\caption{Volume fraction of $\beta$-phase as a function of temperature for several values of hydrogen concentrations in Zircaloy-4  calculated by using model B at heating rates: (a) 10 K/s and (b) 100 K/s.}
\label{fig:BetaHeat-htcalc}
\end{figure}

In order to get an overview of the influence of hydrogen on $\alpha/\beta$ phase transformation in Zircaloy, we have shown the results of our model B computations in Fig. \ref{fig:BetaHeat-htcalc}. The plots in this figure display the volume fraction of $\beta$ phase as a function of temperature for several values of hydrogen concentration in Zircaloy-4  at heating rates of 10 K/s and 100 K/s. It is seen that as hydrogen content in the material is increased, the transition from $\alpha$ to $\beta$ phase initiates at a lower temperature.

\begin{figure}[t]
\centering
\subfigure[]{\label{fig:beta10-uncert}\includegraphics[scale=0.80]{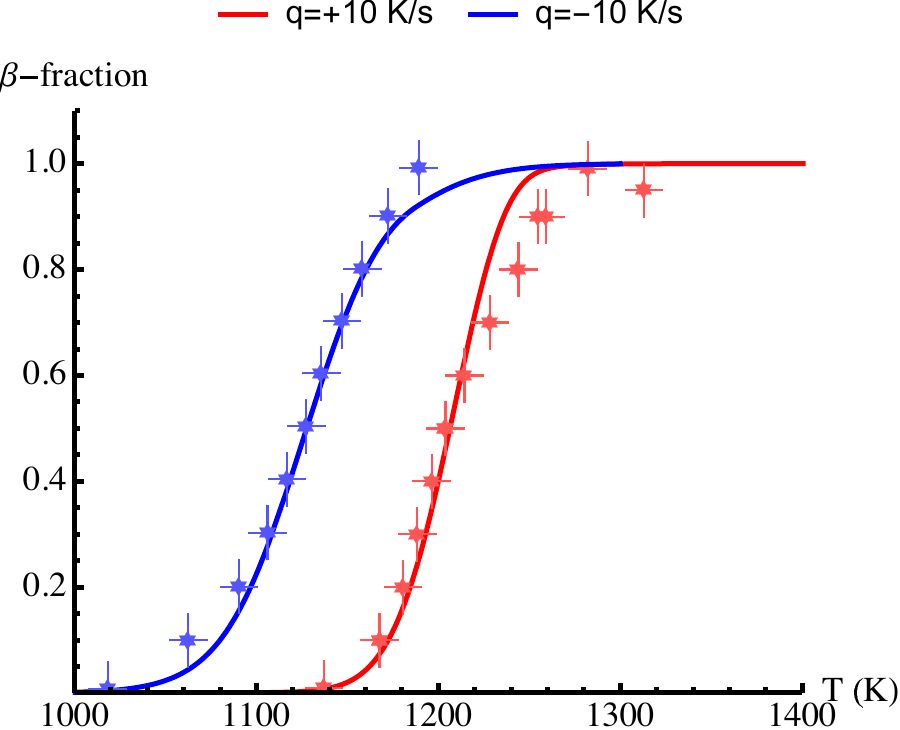}}
\subfigure[]{\label{fig:beta100-uncert}\includegraphics[scale=0.80]{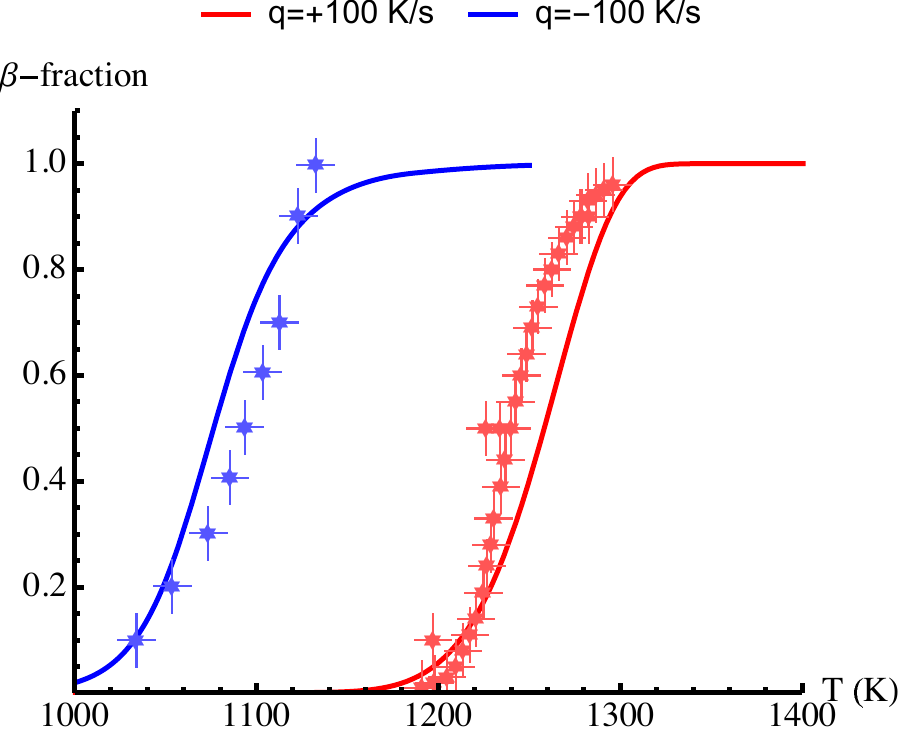}}
\caption{Volume fraction of $\beta$ phase as a function of temperature calculated (lines) by using model B vs. measured data (markers) for Zircaloy-4. The bars indicate $\pm$0.05 and $\pm10$ K uncertainties on measured $\beta$-fraction and temperature, respectively. (a) $q=dT/dt= \pm$10 K/s and (b) $q = \pm$100 K/s.}
\label{fig:BetaHeat-uncert}
\end{figure}

As has been noted the deviation between model B computations and measurements are most prominent for the case of Zircaloy-4 subjected to a heating rate of 100 K/s (cf. Fig. \ref{fig:beta-heat-loj2}) and even, to some extend, at a cooling rate of $-100$ K/s, cf. Fig. \ref{fig:beta-cool-loj2}. Thereupon, an uncertainty evaluation of measured data and their deviation from the model calculation are worthy of estimation. As alluded in section \ref{sec:expdata}, the uncertainty in measured volume fraction of $\beta$ phase as a function of temperature is estimated to be about $\sigma=\pm 0.05$. Moreover, there is an uncertainty  on temperature measurements of around $\pm 10$ K. In Figs. \ref{fig:beta10-uncert} and \ref{fig:beta100-uncert}, we have plotted the measured data for Zircaloy-4 for $q=dT/dt= \pm$10 K/s and (b) $q=dT/dt= \pm$100 K/s including  uncertainties (bars) of $\sigma=\pm 0.05$ on  $\beta$ phase volume fraction and $\pm 10$ K on temperature. We have also included in these figures the corresponding model B computations for comparison.As can be seen the model B computations for the case of $q=\pm$10 K/s are reasonably satisfactory. However, for the case of $q=\pm$100 K/s, the model B deviates from measured $\beta$ phase fractions up to around 0.15 in certain temperature ranges, Fig. \ref{fig:beta100-uncert}. For Zr1NbO at heating rates $q=+10$ and +100 K/s, Fig. \ref{fig:beta-heat-m5}, the data are reproduced fairly satisfactory.

\section{Discussion}
\label{sec:discuss}
The phase boundary temperatures versus excess oxygen concentration, calculated through Eq. (\ref{eq:T_phases}), were compared with experimental data for Zircaloy-4 and Zr-0.5\%NbO cladding in Fig. \ref{fig:zry-o}. The  data were obtained from metallographic analyses of cladding tube samples, quenched from different equilibrium temperatures. We should note that $T_{\beta}$, calculated through Eq. (\ref{eq:T_phases}), may exceed the Zircaloy or Zr1NbO melting temperatures at high oxygen concentrations. The melting (solidus) temperature of Zircaloy-4 increases from about 2025 K at zero excess oxygen concentration to about 2320 K for $x_\mathrm{O} \ge$ 0.04 \cite{Hayward_George_1999}.

\begin{figure}[!hbtp]
  \begin{center}
    \includegraphics[width=0.50\textwidth]{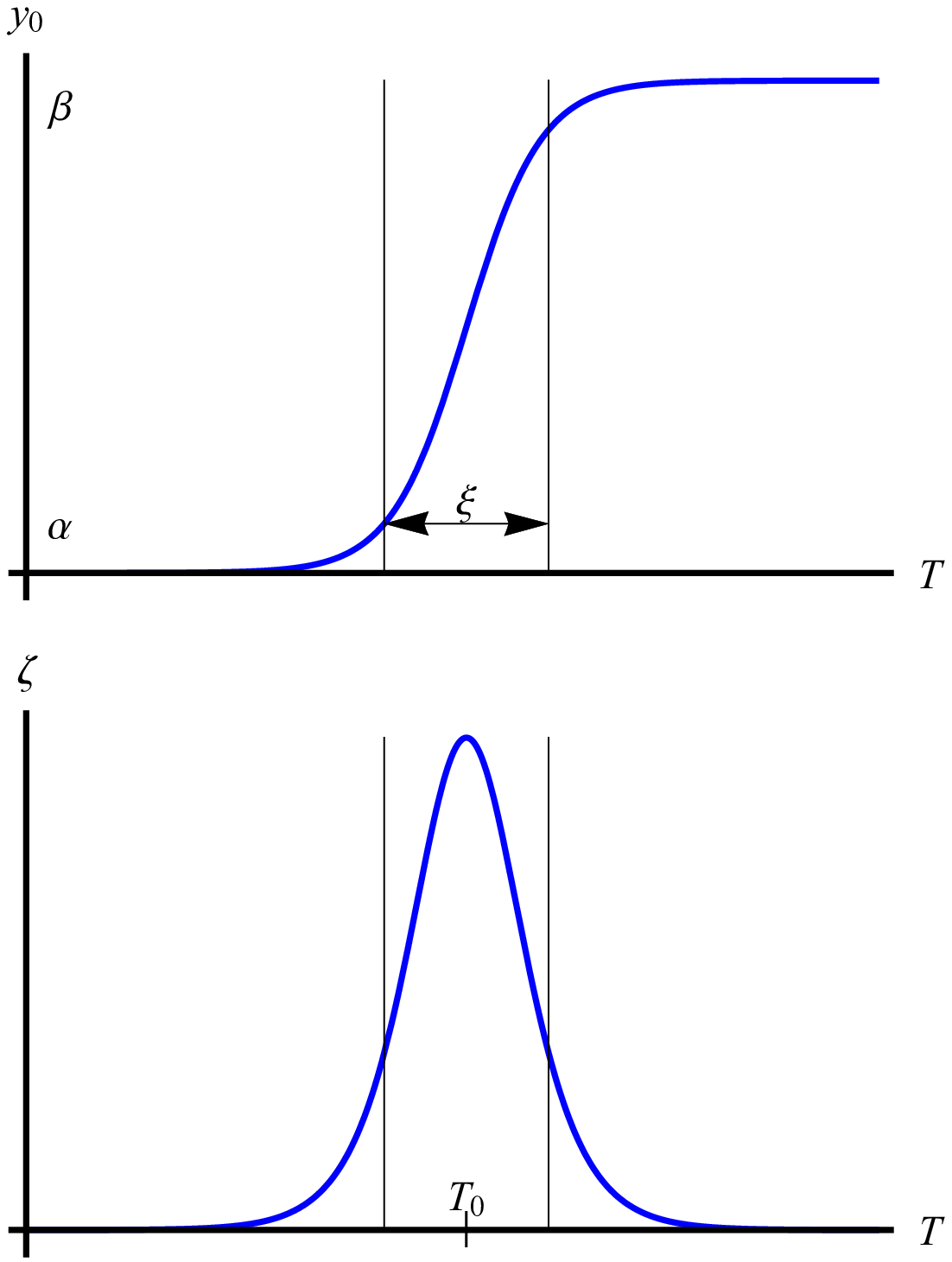}
    \caption{Schematic plots of $\alpha/\beta$ phase transition in equilibrium. \textbf{Top}: the hyperbolic tangent profile of $y_0$ corresponding to two coexisting phases in the span of $\xi$. \textbf{Bottom}: The slope of the profile $\zeta=dy_0/dT$ indicating the sharpness of the transition.}
    \label{fig:soliton-ab}
  \end{center}
\end{figure}

Turning now to the thermal equilibrium relation \eqref{eq:PC_equ}, we stated this formula without any theoretical justification or explanation, albeit that it captures experimental data adequately. Let's write the hyperbolic tangent part of this equation (the first term 1/2 is just a constant shift on the ordinate) in a simplified form
\begin{equation}
\label{eq:soliton}
y_0(T) = \frac{1}{2} \tanh{\left( \frac{T-T_0}{\sqrt{2}\xi} \right)},
\end{equation}
where  $T_0 \equiv T_\mathrm{mid}$ and $\xi   \equiv T_\mathrm{span}/\sqrt 2$. Indeed, Eq. \eqref{eq:soliton} is a stationary solution of the Cahn-Hilliard equation of phase field theory in the temperature domain, namely
\begin{equation}
\label{eq:cahn-hilliard-static}
\xi^2 \frac{d^2 y_0}{dT^2} = \frac{\delta f[y_0(T)]}{\delta y_0(T)},
\end{equation}
where $f(y)=y^4-y^2/2$ is a double-well function (potential) of $y$; see e.g. \cite{Chaikin_Lubensky_1995,Desai_Kapral_2009}. This can be verified either by direct substitution or by integrating Eq. \eqref{eq:cahn-hilliard-static} with appropriate  boundary conditions, e.g.  $y_0(T_0)=0$ and $y_0^\prime(T_0)=\sqrt{2}/(4\xi)$. Differentiation of Eq. \eqref{eq:soliton} with respect to $T$ gives
\begin{equation}
\label{eq:soliton-wave}
\zeta(T) \equiv y_0^\prime(T)=\frac{1}{2\sqrt2\xi} \mathrm{sech}^2{\left( \frac{T-T_0}{\sqrt{2}\xi} \right)}.
\end{equation}
It is seen that the parameter $\xi$ appears both in the argument of the sech-function and as its coefficient in inverse. In phase field theory parlance, $\xi$ is the \emph{correlation length}, here  the  temperature span of the coexisting phase. It plays an important role in the analysis of phase transitions. Figure \ref{fig:soliton-ab} shows schematic plots of $y_0$ and $\zeta$ as a function of temperature. It is seen that  $\zeta$  is highly localized in temperature with a maximum at the center of the bump, $T=T_0$, and falling rapidly to zero for $|T-T_0|>\sqrt2\xi$.

Experimental data indicate that the  level of impurities in the material affects the width of $\xi$. For example, as hydrogen concentration in Zircaloy-4 is increased, so does $\xi$, namely at $\bar x_\mathrm{H}\approx 0$ wppm, $\xi\approx 27$ K, whereas at  $\bar x_\mathrm{H} = 500$ wppm,  $\xi\approx 37$ K; using  relation \eqref{eq:T_s} and data in table   \ref{tab:Phase_boundaries} for Zircaloy-4 with $\xi   \equiv T_\mathrm{span}/\sqrt 2$. In regard to the influence of oxygen on the Zr-O phase equilibrium, oxygen is an "$\alpha$ stabilizer". It expands the $\alpha$ phase region of the phase diagram by formation of an interstitial solid solution in zirconium \cite{Lemaignan_Motta_1994}. That is, excess oxygen widens $\xi$. For example, at an excess oxygen concentration of 1wt\%, $\xi\approx 80$ K; see  relation \eqref{eq:T_s} and data in table   \ref{tab:Phase_boundaries} for Zircaloy-4. We may also compare $\xi$ for Zr1NbO vs. Zircaloy-4. Again using data in table  \ref{tab:Phase_boundaries}, we see that for $x_\mathrm{H}\approx 0$ wppm, $\xi_\mathrm{Zr1NbO}\approx 34$ K, i.e. somewhat wider than that for Zircaloy-4.  This is expected, since Nb is a $\beta$ stabilizer, meaning that it widens the $(\alpha+\beta$) domain in Zr-based alloys \cite{Lemaignan_Motta_1994}.

\begin{figure}[!hbtp]
  \begin{center}
    \includegraphics[width=0.75\textwidth]{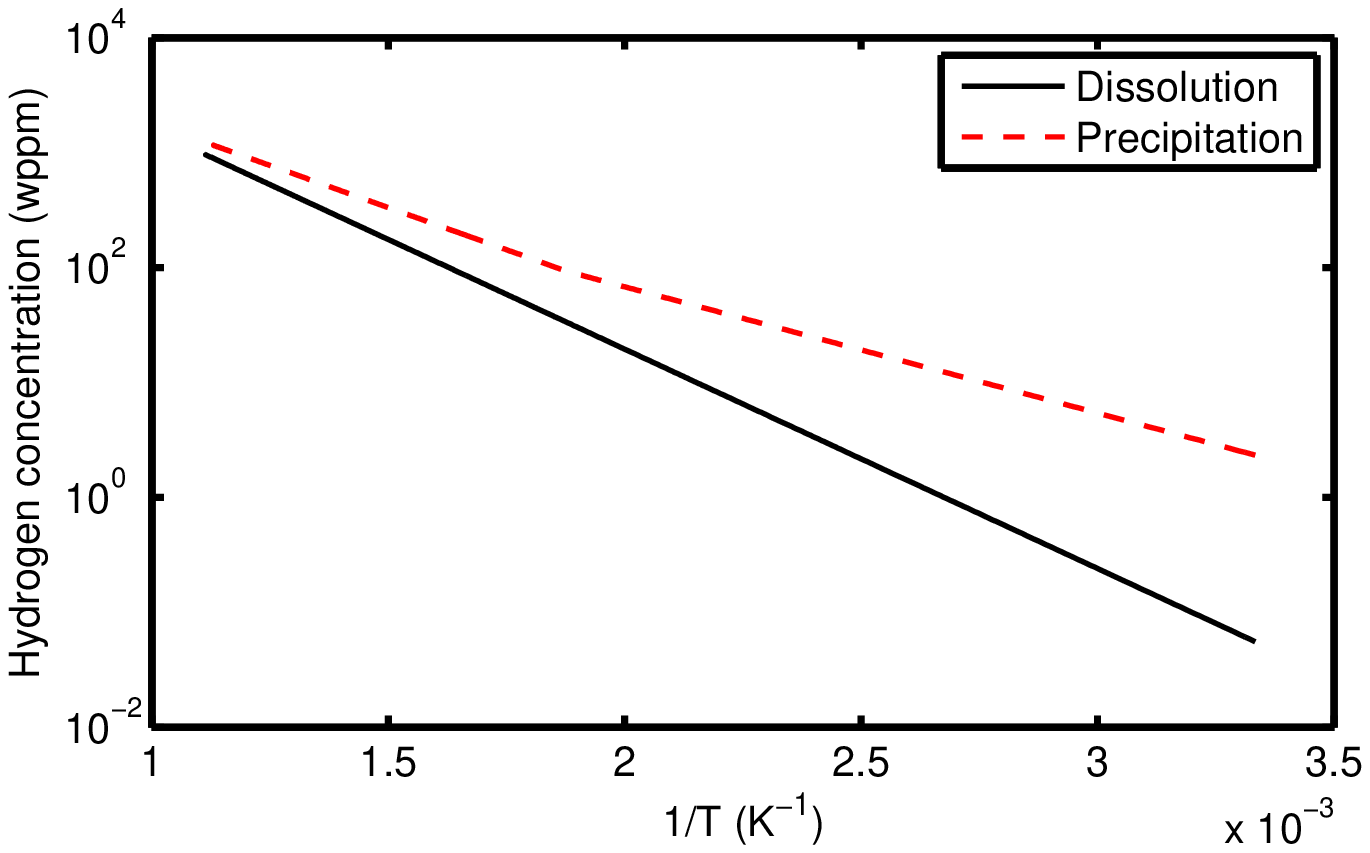}
    \caption{Solid solubility limits for hydrogen in Zircaloy versus inverse temperature, i.e. hydride dissolution and precipitation. The Arrhenius plots are based on measurements of Une and Ishimoto \cite{une2003dissolution} in the temperature range of about 320 to 900 K.}
    \label{fig:fig/tss_une06}
  \end{center}
\end{figure}

Regarding the issue of hydrogen in Zircaloy,  we should say that in the studied range of temperature and hydrogen concentration, cf. figures  \ref{fig:zry4-eqh}, \ref{fig:BetaHeat_htField}, \ref{fig:BetaHeat-htcalc}, hydrogen is primarily in solid solution, i.e. not in zirconium hydride phases. This can be checked from data on the temperature dependence of hydrogen solubility limit in zirconium alloys \cite{yamanaka1997study,une2003dissolution} and the ensuing phase diagram \cite{yamanaka1997study}. It is, however, worth recalling that the temperature dependence of the hydrogen solubility limits for hydride dissolution (during heat-up) and precipitation (during cool-down) are different in Zircaloys \cite{une2003dissolution}; see Fig. \ref{fig:fig/tss_une06}. 

Turning now into the issues in kinetics, the $\alpha\leftrightharpoons\beta$ phase transformation in zirconium base alloys comprise the process of nucleation and growth. For instance, on heating from $\alpha$ phase, the transformation involves $\beta$-nucleation, $\beta$-growth and $\beta$-completion and vice versa; i.e. $\alpha$-nucleation, $\alpha$-growth and $\alpha$-completion. A commonly used macroscopic model for this process is the Kolmogorov-Johnson-Mehl-Avrami \cite{Kolmogorov_1991b,Johnson_Mehl_1939,Avrami_1939,Avrami_1940} or KJMA description \cite{Sekimoto_1986,Onuki_2002}. The KJMA model assumes that tiny spherical grains nucleate at a constant rate per unit volume in the metastable phase and grow isotropically at constant velocity once formed. As first formulated by Kolmogorov \cite{Kolmogorov_1991b}, the volume fraction of the favored phase at time $t$ can be expressed as
\begin{equation}
\label{eq:kjma}
\phi(t)=1-\exp\Big[-\frac{\pi}{3}\Big(\frac{t}{\tau_d}\Big)^{d+1}\Big],
\end{equation}
where $d$ is the space dimension, $\tau_d=(Iv^d)^{-1/(d+1)}$ is a characteristic time, $I$ is the nucleation rate per unit volume with dimension unit $[\mathrm{T^{-1}L^{-d}}]$ and $v$ is the interface velocity (dimension [$\mathrm{LT^{-1}}$]). In differential form, it reads
\begin{equation}
\label{eq:kjma-dif}
\frac{d\phi}{dt}=\frac{\pi}{3}(d+1)IR_m^d(1-\phi).
\end{equation}
where $R_m \equiv vt$ is the grain radius obeying an algebraic growth law. The nucleation process in  Kolmogorov's theory is considered as homogeneous, meaning that, it occurs uniformly throughout the metastable phase. In most cases, however, nucleation is heterogeneous. That is, nuclei primarily form  at impurities and defects in the crystal, which are present before the transformation started. A simple  model in the spirit of KJMA has been discussed in \cite{Bradley_Strenski_1989}, which places nuclei randomly with density $\rho$ throughout the solid. When the phase transformation is initiated, spherical grains form at each nuclei and grow isotropically with velocity $v$.  Both the KJMA model and its simplified heterogeneous nucleation variant \cite{Bradley_Strenski_1989} are a mean-field type theory. Meaning that, there is no spatial variation  in $\phi$ through the solid, i.e. $\nabla\phi=0$.

In \cite{Massih_Jernkvist_2009}, we made a detailed comparison between Eq. \eqref{eq:kjma} and our model A. To sum up that account, we note that under isothermal condition, by putting $\phi=y/y_\mathrm{eq}$, we can write Eq. \eqref{eq:moda-kin} in the form
\begin{equation}
\label{eq:moda-kin-norm}
\frac{d\phi}{dt}  = \frac{1}{\tau}\big(1-\phi\big).
\end{equation}
Similarly, the model B kinetics, as described by Eq. \eqref{eq:modb-kin-hc}, can be written as 
\begin{equation}
\label{eq:modb-kin-norm}
\frac{d\phi}{dt}  = \frac{1}{\tau}\Big(\big(1-\phi\big) + b \big(1-\phi\big)^2\Big)
\end{equation}
Now a naive comparison of Eq. \eqref{eq:kjma-dif} with \eqref{eq:moda-kin-norm} yields
\begin{equation}
\label{eq:tau-tau}
\tau^{-1} = \frac{\pi}{3}(d+1) R_m^d I.
\end{equation}
That is, the relaxation time used in our kinetic model is inversely proportional  to the nucleation rate $I$ and $R_m^d$, cf. Eq. \eqref{eq:chartime-2}, $R_m^dI \sim \exp[-E/T(t)]/B_i$, which its time dependence is merely through temperature. In the material science literature, Eq. \eqref{eq:kjma} is often expressed in the simple form \cite{Porter_Easterling_1981}
\begin{equation}
\label{eq:avrami}
\phi(t)=1-\exp[- K t^n],
\end{equation}
where $K$ is a function of temperature designating the nucleation and growth rates, and $n$ is a numerical exponent (Avrami exponent) whose value may vary from $\sim 1$ to 4. According to  \cite{Porter_Easterling_1981} if there is no change in the nucleation mechanism, $n$ is independent of temperature. So, as can be inferred from Eq. \eqref{eq:moda-kin-norm}, our model A corresponds to $n=1$ and $K=1/\tau$.

\begin{figure}[!hbtp]
  \begin{center}
    \includegraphics[width=0.60\textwidth]{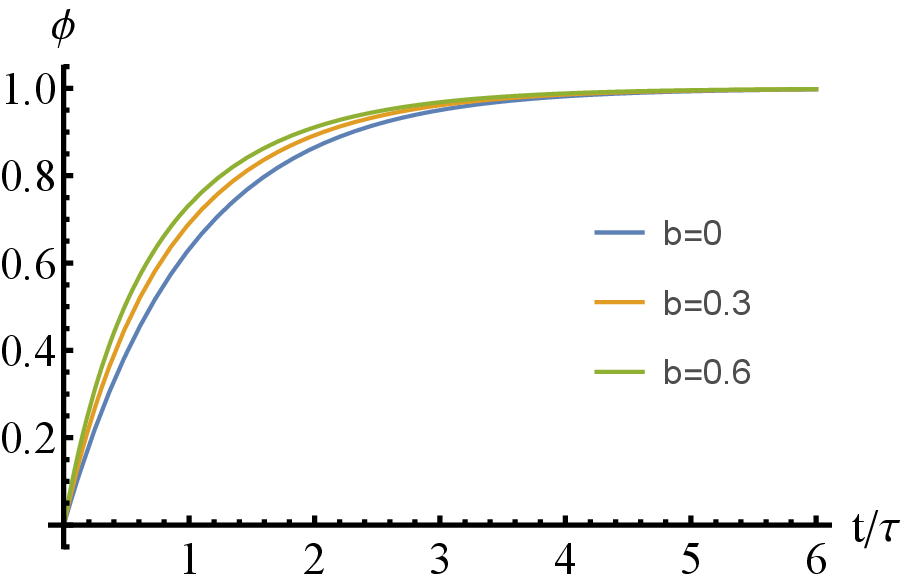}
    \caption{Plots of the relation \eqref{eq:loj2_analytic} for several values of the model parameter $b$. Model A corresponds to $b=0$, while for model B, we have chosen $b=0.3$, cf. table \ref{tab:relax-time}.}
    \label{fig:loj-analytic}
  \end{center}
\end{figure}

The behavior of the model B, as described by Eq. \eqref{eq:modb-kin-norm}, is slightly more complicated. Solution to Eq. \eqref{eq:modb-kin-norm} subject to $\phi(0)=0$ under isothermal condition reads
\begin{equation}
\label{eq:loj2_analytic}
\phi(t)=1-\frac{e^{-t/\tau}}{1+b(1-e^{-t/\tau})},
\end{equation}
which is not in the form of the Avrami relation \eqref{eq:avrami}, except for $b=0$. For the sake of illustration, we have plotted this solution in Fig. \ref{fig:loj-analytic} for several values of $b$. As can be seen from these plots, the behavior for $b=0$  and $b=0.3$ is rather close. So, we expect even for $b=0.3$ the Avrami exponent should be around one.

It is worth mentioning here that the space dimension $d$ (or $n\equiv d+1$) appearing in the foregoing equations depends on genre of nucleation and/or growth. Avrami \cite{Avrami_1940} attributes $n=1$ to 1-d growth (needles), $n=2$ to 2-d growth (plates), $n=3$ to 3-d growth (spheres or polyhedra),  and $n=4$ to nucleation and 3-d growth.\cite{rubie1985kinetics} Whereas Cahn, \cite{Cahn_1956a} in a subsequent theoretical study of the Avrami model, relations of type \eqref{eq:avrami}, concludes that after site saturation, the stage in a reaction at which all potential nucleation sites are consumed,  the reaction proceeds by growth alone. In Cahn's analysis $n=1$ describes nucleation on grain surfaces, $n=2$ nucleation on grain edges and $n=3$ nucleation on grain corners.\cite{rubie1985kinetics}

We should note that our model B is tailored for fuel rod behavior computer program application, e.g. the \texttt{FRAPTRAN} program \cite{Geelhood_et_al_2014,Jernkvist_Massih_2019}, where its parameters are independent of thermal rate $dT/dt$, contrary to model A. Its performance against experimental data is fairly good, considering the uncertainty in the measured measured phase fraction and temperature, which is about $\pm 0.05$ and $\pm 10$ K, respectively. The model, despite some theoretical justification, is basically empirically based with adjustable parameter constants. Its results perhaps can be improved further by more tune-up of those  constants. The model should be applicable to thermal rating in the range $dT/dt=-100$ to +100 K/s and hydrogen contents in the alloys up to around 1000 wppm; however, with more uncertainty toward  higher thermal rates.

\section{Summary and conclusions}
\label{sec:conclude}
In this paper we have introduced a new model (model B) for the kinetics of $\alpha \rightleftharpoons \beta$ transformation in zirconium alloys used in the nuclear industry as fuel cladding material. We first surveyed pertinent experimental data for the alloys reported in the literature, section \ref{sec:expdata}. They constitute  an empirical basis for our model. We then presented our model B together with our previously published  model A, the latter for the sake of comparison and completeness, section \ref{sec:models}. 
The main difference between models A and B is that model B is formulated such that it can be applied to arbitrary heating/cooling histories, while model A is restricted to cases with a constant heating/cooling rate. Moreover, the kinetic equation for model B extends to a higher order term in Taylor expansion about the equilibrium. Both models are extended to account for the effect of excess oxygen and hydrogen content in the alloy. Surplus in oxygen  may occur under an accident situation due to the high temperature metal-water reaction as the cladding temperature raises say from roughly 650 K to 1470 K. The increase in the hydrogen content of cladding from its as-fabricated value of around 10 wppm occurs during normal reactor operation and is enhanced during an accident, again due to the metal-water reaction. All the experiments which we have evaluated with our models were performed on unirradiated materials. The effect of irradiation on $\alpha \rightleftharpoons \beta$ is believed not to be significant, at least for the rather low heating rates expected in light water reactor LOCA; cf. Fig. \ref{fig:QL1}. The thermodynamic stability of crystal structures is the dominant factor for phase transformation. During $\alpha \to \beta$ phase transformation as the cladding temperature raises, the irradiation-induced defects would mostly be removed as the recrystallizing front is swept through the lattice \cite{hindle1983annealing}.

We compared the results of our model computations with experimental data as regards the relative amounts of $\beta$ (or $\alpha$) phase fractions as a function of temperature in the alloys, section \ref{sec:compute}. Our computations are mainly on Zircaloy-4, including hydrogenated material, and to a limited degree on Zr1NbO material because of dearth of available measurements for this alloy. The agreement between calculations and measurements is fairly satisfactory considering the complexity in the details and the measurement uncertainties. Finally, in section \ref{sec:discuss} we made a theoretical analysis of the thermal equilibrium expression used and qualified for the transformation. In that section, we also briefly discussed the kinetics in regard to nucleation and growth, and delineated  the range of applicability of the putative model.

Our model B is basically empirically based, with adjustable parameter constants. It  is intended for implementation in fuel rod behavior computer programs and reproduces available experimental data fairly satisfactory.

\section*{Acknowledgements}
We are thankful to Tatiana Taurines for pointing out the shortcomings in an earlier version of model B and providing us ref. \cite{Frechinet_2001} and Jean-Christophe Brachet  for helpful communications regarding measurements in ref. \cite{Forgeron_et_al_2000}.
The work was partially supported by the Swedish Radiation Safety Authority (SSM) under the contract number SSM2018-4296.

\section*{Author contributions statement}

A.R.M. and L.O.J. contributed equally to the content of this work.  All authors reviewed the manuscript. 

\section*{Additional information}
\textbf{Competing financial interests:} The authors declare no competing financial interests.


\begin{thebibliography}{10}
\urlstyle{rm}
\expandafter\ifx\csname url\endcsname\relax
  \def\url#1{\texttt{#1}}\fi
\expandafter\ifx\csname urlprefix\endcsname\relax\def\urlprefix{URL }\fi
\expandafter\ifx\csname doiprefix\endcsname\relax\def\doiprefix{DOI: }\fi
\providecommand{\bibinfo}[2]{#2}
\providecommand{\eprint}[2][]{\url{#2}}

\bibitem{Lutjering_Williams_2007}
\bibinfo{author}{L\"utjering, G.} \& \bibinfo{author}{Williams, J.~C.}
\newblock \emph{\bibinfo{title}{Titanium}} (\bibinfo{publisher}{Springer},
  \bibinfo{address}{Heidelberg}, \bibinfo{year}{2007}),
  \bibinfo{edition}{second} edn.

\bibitem{Boyer_2010}
\bibinfo{author}{Boyer, R.~R.}
\newblock \bibinfo{journal}{\bibinfo{title}{Attributes, characteristics, and
  applications of titanium and its alloys}}.
\newblock {\emph{\JournalTitle{JOM}}} \textbf{\bibinfo{volume}{65}},
  \bibinfo{pages}{21--24} (\bibinfo{year}{2010}).

\bibitem{zhang2019review}
\bibinfo{author}{Zhang, L.-C.} \& \bibinfo{author}{Chen, L.-Y.}
\newblock \bibinfo{journal}{\bibinfo{title}{A review on biomedical titanium
  alloys: recent progress and prospect}}.
\newblock {\emph{\JournalTitle{Advanced Engineering Materials}}}
  \textbf{\bibinfo{volume}{21}}, \bibinfo{pages}{1801215}
  (\bibinfo{year}{2019}).

\bibitem{Lemaignan_Motta_1994}
\bibinfo{author}{Lemaignan, C.} \& \bibinfo{author}{Motta, A.~T.}
\newblock \bibinfo{title}{Zirconium alloys in nuclear applications}.
\newblock In \bibinfo{editor}{Cahn, R.~W.}, \bibinfo{editor}{Haasen, P.} \&
  \bibinfo{editor}{Kramer, E.~J.} (eds.) \emph{\bibinfo{booktitle}{Nuclear
  Materials}}, vol. \bibinfo{volume}{10B} of \emph{\bibinfo{series}{Materials
  Science and Technology}}, chap.~\bibinfo{chapter}{7}, \bibinfo{pages}{1--51}
  (\bibinfo{publisher}{VCH}, \bibinfo{address}{Weinheim, Germany},
  \bibinfo{year}{1994}).
\newblock \bibinfo{note}{Volume editor B.R.T. Frost}.

\bibitem{suzuki2020appraising}
\bibinfo{author}{Suzuki, A.~K.} \emph{et~al.}
\newblock \bibinfo{journal}{\bibinfo{title}{Appraising the potential of
  \mbox{Z}r-based biomedical alloys to reduce magnetic resonance imaging
  artifacts}}.
\newblock {\emph{\JournalTitle{Scientific Reports}}}
  \textbf{\bibinfo{volume}{10}}, \bibinfo{pages}{1--7} (\bibinfo{year}{2020}).

\bibitem{Tricot_1992}
\bibinfo{author}{Tricot, R.}
\newblock \bibinfo{journal}{\bibinfo{title}{The metallurgy and functional
  properties of hafnium}}.
\newblock {\emph{\JournalTitle{J. Nucl. Mater.}}}
  \textbf{\bibinfo{volume}{189}}, \bibinfo{pages}{277--288}
  (\bibinfo{year}{1992}).

\bibitem{Miquet_et_al_1982a}
\bibinfo{author}{Miquet, A.}, \bibinfo{author}{Charquet, D.} \&
  \bibinfo{author}{Allibert, C.~H.}
\newblock \bibinfo{journal}{\bibinfo{title}{Solid state phase equilibria of
  \mbox{Z}ircaloy-4 in the temperature range 750-1050$^\circ$\mbox{C}}}.
\newblock {\emph{\JournalTitle{J. Nucl. Mater.}}}
  \textbf{\bibinfo{volume}{105}}, \bibinfo{pages}{132--141}
  (\bibinfo{year}{1982}).

\bibitem{Perez_Massih_2007}
\bibinfo{author}{Per\'{e}z, R.~J.} \& \bibinfo{author}{Massih, A.~R.}
\newblock \bibinfo{journal}{\bibinfo{title}{Thermodynamic evaluation of the
  \mbox{Nb-O-Zr} system}}.
\newblock {\emph{\JournalTitle{J. Nucl. Mater.}}}
  \textbf{\bibinfo{volume}{360}}, \bibinfo{pages}{242--254}
  (\bibinfo{year}{2007}).

\bibitem{Kaddour_et_al_2004}
\bibinfo{author}{Kaddour, D.} \emph{et~al.}
\newblock \bibinfo{journal}{\bibinfo{title}{Experimental determination of creep
  properties of zirconium alloys together with phase transformation}}.
\newblock {\emph{\JournalTitle{Scripta Materialia}}}
  \textbf{\bibinfo{volume}{51}}, \bibinfo{pages}{515--519}
  (\bibinfo{year}{2004}).

\bibitem{Terai_et_al_1997}
\bibinfo{author}{Terai, T.}, \bibinfo{author}{Takahashi, Y.},
  \bibinfo{author}{Masumura, S.} \& \bibinfo{author}{Yoneoka, T.}
\newblock \bibinfo{journal}{\bibinfo{title}{Heat capacity and phase transition
  of \mbox{Z}ircaloy-4}}.
\newblock {\emph{\JournalTitle{J. Nucl. Mater.}}}
  \textbf{\bibinfo{volume}{247}}, \bibinfo{pages}{222--226}
  (\bibinfo{year}{1997}).

\bibitem{Forgeron_et_al_2000}
\bibinfo{author}{Forgeron, T.} \emph{et~al.}
\newblock \bibinfo{title}{Experiment and modelling of advanced fuel rod
  cladding behavior under \mbox{LOCA} conditions: Alpha-beta phase
  transformation kinetics and \mbox{EDGAR} methodology}.
\newblock In \emph{\bibinfo{booktitle}{Zirconium in the Nuclear Industry:
  Twelfth International Symposium}}, ASTM STP 1354, \bibinfo{pages}{256--278}
  (\bibinfo{organization}{American Society for Testing and Materials},
  \bibinfo{year}{2000}).

\bibitem{Massih_Jernkvist_2009}
\bibinfo{author}{Massih, A.~R.} \& \bibinfo{author}{Jernkvist, L.~O.}
\newblock \bibinfo{journal}{\bibinfo{title}{Transformation kinetics of alloys
  under non-isothermal conditions}}.
\newblock {\emph{\JournalTitle{Modelling Simul. Mater. Sci. Eng.}}}
  \textbf{\bibinfo{volume}{17}}, \bibinfo{pages}{055002}
  (\bibinfo{year}{2009}).

\bibitem{Massih_2009}
\bibinfo{author}{Massih, A.~R.}
\newblock \bibinfo{journal}{\bibinfo{title}{Transformation kinetics of
  zirconium alloys under non-isothermal conditions}}.
\newblock {\emph{\JournalTitle{J. Nucl. Mater.}}}
  \textbf{\bibinfo{volume}{384}}, \bibinfo{pages}{330--335}
  (\bibinfo{year}{2009}).

\bibitem{Chung_Kassner_1979}
\bibinfo{author}{Chung, H.~M.} \& \bibinfo{author}{Kassner, T.~F.}
\newblock \bibinfo{journal}{\bibinfo{title}{Pseudobinary \mbox{Zircaloy}-oxygen
  phase diagram}}.
\newblock {\emph{\JournalTitle{J. Nucl. Mater.}}}
  \textbf{\bibinfo{volume}{84}}, \bibinfo{pages}{327--339}
  (\bibinfo{year}{1979}).

\bibitem{Hunt_Niessen_1970}
\bibinfo{author}{Hunt, C. E.~L.} \& \bibinfo{author}{Niessen, P.}
\newblock \bibinfo{journal}{\bibinfo{title}{The effect of oxygen on the
  equilibrium $\beta/\alpha+\beta$ transformation temperature of
  zirconium-niobium alloys}}.
\newblock {\emph{\JournalTitle{J. Nucl. Mater.}}}
  \textbf{\bibinfo{volume}{35}}, \bibinfo{pages}{134--136}
  (\bibinfo{year}{1970}).

\bibitem{Toffolon_et_al_2002}
\bibinfo{author}{Toffolon, C.} \emph{et~al.}
\newblock \bibinfo{title}{Experimental study and preliminary thermodynamic
  calculations of the pseudo-ternary \mbox{Zr-Nb-Fe-(O,Sn)} system}.
\newblock In \emph{\bibinfo{booktitle}{Zirconium in the Nuclear Industry:
  Twelfth International Symposium}}, ASTM STP 1423, \bibinfo{pages}{361--383}
  (\bibinfo{organization}{American Society for Testing and Materials},
  \bibinfo{year}{2002}).

\bibitem{Corchia_Righini_1981}
\bibinfo{author}{Corchia, M.} \& \bibinfo{author}{Righini, F.}
\newblock \bibinfo{journal}{\bibinfo{title}{Kinetic aspects of the phase
  transformations in \mbox{Z}ircaloy-2}}.
\newblock {\emph{\JournalTitle{J. Nucl. Mater.}}}
  \textbf{\bibinfo{volume}{97}}, \bibinfo{pages}{137--148}
  (\bibinfo{year}{1981}).

\bibitem{Arias_Roberti_1983}
\bibinfo{author}{Arias, D.} \& \bibinfo{author}{Roberti, L.}
\newblock \bibinfo{journal}{\bibinfo{title}{The solubility of tin in $\alpha$
  and $\beta$ zirconium below \mbox{1000$^\circ$C}}}.
\newblock {\emph{\JournalTitle{J. Nucl. Mater.}}}
  \textbf{\bibinfo{volume}{118}}, \bibinfo{pages}{143--149}
  (\bibinfo{year}{1983}).

\bibitem{Arias_Guerra_1987}
\bibinfo{author}{Arias, D.} \& \bibinfo{author}{Guerra, R.~C.}
\newblock \bibinfo{journal}{\bibinfo{title}{Phase transition temperature in
  \mbox{Z}ircaloy-2}}.
\newblock {\emph{\JournalTitle{J. Nucl. Mater.}}}
  \textbf{\bibinfo{volume}{144}}, \bibinfo{pages}{196--199}
  (\bibinfo{year}{1987}).

\bibitem{Brachet_et_al_2002}
\bibinfo{author}{Brachet, J.-C.} \emph{et~al.}
\newblock \bibinfo{title}{Influence of hydrogen content on the $\alpha$/$\beta$
  phase transformation temperatures and on the thermal-mechanical behavior of
  \mbox{Zy-4}, \mbox{M4 (ZrSnFeV)}, and \mbox{M5(ZrNbO)} alloys during the
  first phase of \mbox{LOCA} transient}.
\newblock In \emph{\bibinfo{booktitle}{Zirconium in the Nuclear Industry:
  Thirteenth International Symposium}}, ASTM STP 1423,
  \bibinfo{pages}{673--701} (\bibinfo{organization}{ASTM International},
  \bibinfo{year}{2002}).

\bibitem{Frechinet_2001}
\bibinfo{author}{Frechinet, S.}
\newblock \emph{\bibinfo{title}{Transformations et comportements du
  \mbox{Z}ircaloy-4 en conditions anisothermes}}.
\newblock \bibinfo{type}{Doctoral dissertation}, \bibinfo{school}{Ecoles des
  Mines de Paris}, \bibinfo{address}{Paris, France} (\bibinfo{year}{2001}).

\bibitem{Nguyen_2017}
\bibinfo{author}{Nguyen, C.-T.}
\newblock \emph{\bibinfo{title}{Microstructure changes during fast $\beta$
  cycles of zirconium alloys}}.
\newblock \bibinfo{type}{\mbox{D Phil} thesis}, \bibinfo{school}{Manchester
  University}, \bibinfo{address}{Manchester, UK} (\bibinfo{year}{2017}).

\bibitem{Nguyen_et_al_2017}
\bibinfo{author}{Nguyen, C.-T.}, \bibinfo{author}{Romero, J.},
  \bibinfo{author}{Ambard, A.}, \bibinfo{author}{Preuss, M.} \&
  \bibinfo{author}{da~Fonseca, J.~Q.}
\newblock \bibinfo{title}{Predicting the flow stress of \mbox{Zircaloy-4} under
  in-reactor accident conditions}.
\newblock In \emph{\bibinfo{booktitle}{Zirconium in the Nuclear Industry:
  Eighteenth International Symposium}}, ASTM STP 1597,
  \bibinfo{pages}{214--239} (\bibinfo{organization}{American Society for
  Testing and Materials}, \bibinfo{year}{2017}).

\bibitem{Nguyen_et_al_2019}
\bibinfo{author}{Nguyen, C.-T.}, \bibinfo{author}{Romero, J.},
  \bibinfo{author}{Ambard, A.}, \bibinfo{author}{Preuss, M.} \&
  \bibinfo{author}{da~Fonseca, J.~Q.}
\newblock \bibinfo{journal}{\bibinfo{title}{The effect of cold work on the
  transformation kinetics and texture of a zirconium alloy during fast thermal
  cycling.}}
\newblock {\emph{\JournalTitle{Mater. Sci. \& Eng. A}}}
  \textbf{\bibinfo{volume}{746}}, \bibinfo{pages}{424--433}
  (\bibinfo{year}{2019}).

\bibitem{Jailin_et_al_2020}
\bibinfo{author}{Jailin, T.} \emph{et~al.}
\newblock \bibinfo{journal}{\bibinfo{title}{Experimental study and modelling of
  the phase transformation of \mbox{Z}ircaloy-4 alloy under high thermal
  transients}}.
\newblock {\emph{\JournalTitle{Mater. Character.}}}
  \textbf{\bibinfo{volume}{162}}, \bibinfo{pages}{110199}
  (\bibinfo{year}{2020}).

\bibitem{Benes_et_al_2011}
\bibinfo{author}{B\v{e}nes, O.} \emph{et~al.}
\newblock \bibinfo{journal}{\bibinfo{title}{Kinetic studies of the
  $\alpha-\beta$ phase transition in the \mbox{Zr1\%Nb} cladding for nuclear
  reactors}}.
\newblock {\emph{\JournalTitle{J. Nucl. Mater.}}}
  \textbf{\bibinfo{volume}{414}}, \bibinfo{pages}{88--91}
  (\bibinfo{year}{2011}).

\bibitem{PM18-001v1}
\bibinfo{author}{Jernkvist, L.~O.} \& \bibinfo{author}{Massih, A.~R.}
\newblock \bibinfo{title}{A model for \mbox{Z}r-alloy cladding solid-to-solid
  phase transformation in \texttt{FRAPTRAN-1.4}}.
\newblock \bibinfo{type}{Technical Memorandum} \bibinfo{number}{PM18-001v1},
  \bibinfo{institution}{Quantum Technologies}, \bibinfo{address}{Uppsala,
  Sweden} (\bibinfo{year}{2017}).

\bibitem{Leblond_Devaux_1984}
\bibinfo{author}{Leblond, J.~B.} \& \bibinfo{author}{Devaux, J.}
\newblock \bibinfo{journal}{\bibinfo{title}{A new kinetic model for
  anisothermal metallurgical transformations in steels including effect of
  austenite grain size}}.
\newblock {\emph{\JournalTitle{Acta Metall.}}} \textbf{\bibinfo{volume}{26}},
  \bibinfo{pages}{137--146} (\bibinfo{year}{1984}).

\bibitem{NEA_LOCA_SOAR_2009}
\bibinfo{title}{Nuclear fuel behaviour in loss-of-coolant accident \mbox{LOCA}
  conditions}.
\newblock \bibinfo{type}{State-of-the-art report} \bibinfo{number}{NEA No.
  6846}, \bibinfo{institution}{OECD Nuclear Energy Agency},
  \bibinfo{address}{Paris, France} (\bibinfo{year}{2009}).

\bibitem{Stuckert_et_al_2020}
\bibinfo{author}{Stuckert, J.}, \bibinfo{author}{Grosse, M.},
  \bibinfo{author}{Steinbrueck, M.}, \bibinfo{author}{Walter, M.} \&
  \bibinfo{author}{Wensauer, A.}
\newblock \bibinfo{journal}{\bibinfo{title}{Results of the
  \mbox{QUENCH}-\mbox{LOCA} experimental program at \mbox{KIT}}}.
\newblock {\emph{\JournalTitle{J. Nucl. Mater.}}}
  \textbf{\bibinfo{volume}{534}}, \bibinfo{pages}{152143}
  (\bibinfo{year}{2020}).

\bibitem{Stuckert_et_al_2018}
\bibinfo{author}{Stuckert, J.}, \bibinfo{author}{Grosse, M.},
  \bibinfo{author}{R\"{o}ssger, C.}, \bibinfo{author}{Steinbr\"{u}ck, M.} \&
  \bibinfo{author}{Walter, M.}
\newblock \bibinfo{title}{Results of the \mbox{LOCA} reference bundle test
  \mbox{QUENCH-L1} with \mbox{Zircaloy-4} claddings}.
\newblock \bibinfo{type}{Tech. Rep.} \bibinfo{number}{SR-7651},
  \bibinfo{institution}{Karlsruher Institut f\"{u}r Technologie},
  \bibinfo{address}{Karlsruhe, Germany} (\bibinfo{year}{2018}).

\bibitem{Leistikow_Schanz_1987}
\bibinfo{author}{Leistikow, S.} \& \bibinfo{author}{Schanz, G.}
\newblock \bibinfo{journal}{\bibinfo{title}{Oxidation kinetics and related
  phenomena of \mbox{Zircaloy-4} fuel cladding exposed to high temperature
  steam and hydrogen-steam mixtures under \mbox{PWR} accident conditions}}.
\newblock {\emph{\JournalTitle{Nuclear Engineering and Design}}}
  \textbf{\bibinfo{volume}{103}}, \bibinfo{pages}{65--84}
  (\bibinfo{year}{1987}).

\bibitem{chuto2009high}
\bibinfo{author}{Chuto, T.}, \bibinfo{author}{Nagase, F.} \&
  \bibinfo{author}{Fuketa, T.}
\newblock \bibinfo{journal}{\bibinfo{title}{High temperature oxidation of
  \mbox{Nb}-containing \mbox{Zr} alloy cladding in \mbox{LOCA} conditions}}.
\newblock {\emph{\JournalTitle{Nuclear Eng. Techn.}}}
  \textbf{\bibinfo{volume}{41}}, \bibinfo{pages}{163--170}
  (\bibinfo{year}{2009}).

\bibitem{steinbruck2011oxidation}
\bibinfo{author}{Steinbr{\"u}ck, M.}, \bibinfo{author}{V{\'e}r, N.} \&
  \bibinfo{author}{Gro{\ss}e, M.}
\newblock \bibinfo{journal}{\bibinfo{title}{Oxidation of advanced zirconium
  cladding alloys in steam at temperatures in the range of 600--1200
  \mbox{$^\circ$C}}}.
\newblock {\emph{\JournalTitle{Oxidation of metals}}}
  \textbf{\bibinfo{volume}{76}}, \bibinfo{pages}{215--232}
  (\bibinfo{year}{2011}).

\bibitem{Hagi_Hayashi_1987}
\bibinfo{author}{Hagi, H.} \& \bibinfo{author}{Hayashi, Y.}
\newblock \bibinfo{journal}{\bibinfo{title}{Effects of interstitial impurities
  on dislocation trapping of hydrogen in iron}}.
\newblock {\emph{\JournalTitle{Trans. Japan Inst. Metals}}}
  \textbf{\bibinfo{volume}{28}}, \bibinfo{pages}{375--382}
  (\bibinfo{year}{1987}).

\bibitem{Hayward_George_1999}
\bibinfo{author}{Hayward, P.~J.} \& \bibinfo{author}{George, I.~M.}
\newblock \bibinfo{journal}{\bibinfo{title}{Determination of the solidus
  temperatures of \mbox{Zircaloy-4}/oxygen alloys}}.
\newblock {\emph{\JournalTitle{J. Nucl. Mater.}}}
  \textbf{\bibinfo{volume}{273}}, \bibinfo{pages}{294--301}
  (\bibinfo{year}{1999}).

\bibitem{Chaikin_Lubensky_1995}
\bibinfo{author}{Chaikin, P.~M.} \& \bibinfo{author}{Lubensky, T.~C.}
\newblock \emph{\bibinfo{title}{Principles of Condensed Matter Physics}}
  (\bibinfo{publisher}{Cambridge University Press},
  \bibinfo{address}{Cambridge, UK}, \bibinfo{year}{1995}).
\newblock \bibinfo{note}{Chap. 10}.

\bibitem{Desai_Kapral_2009}
\bibinfo{author}{Desai, R.~C.} \& \bibinfo{author}{Kapral, R.}
\newblock \emph{\bibinfo{title}{Dynamics of Self-Organized and Self-Assembled
  Structures}} (\bibinfo{publisher}{Cambridge University Press},
  \bibinfo{address}{Cambridge, UK}, \bibinfo{year}{2009}).
\newblock \bibinfo{note}{Chap. 7}.

\bibitem{yamanaka1997study}
\bibinfo{author}{Yamanaka, S.}, \bibinfo{author}{Miyake, M.} \&
  \bibinfo{author}{Katsura, M.}
\newblock \bibinfo{journal}{\bibinfo{title}{Study on the hydrogen solubility in
  zirconium alloys}}.
\newblock {\emph{\JournalTitle{J. Nucl. Mater.}}}
  \textbf{\bibinfo{volume}{247}}, \bibinfo{pages}{315--321}
  (\bibinfo{year}{1997}).

\bibitem{une2003dissolution}
\bibinfo{author}{Une, K.} \& \bibinfo{author}{Ishimoto, S.}
\newblock \bibinfo{journal}{\bibinfo{title}{Dissolution and precipitation
  behavior of hydrides in \mbox{Zircaloy-2} and high \mbox{Fe Zircaloy}}}.
\newblock {\emph{\JournalTitle{J. Nucl. Mater.}}}
  \textbf{\bibinfo{volume}{322}}, \bibinfo{pages}{66--72}
  (\bibinfo{year}{2003}).

\bibitem{Kolmogorov_1991b}
\bibinfo{author}{Kolmogorov, A.~N.}
\newblock \bibinfo{title}{On statistical theory of metal crystallization}.
\newblock In \bibinfo{editor}{Shiryayev, A.~N.} (ed.)
  \emph{\bibinfo{booktitle}{Selected Works of A. N. Kolmogorov}},
  vol.~\bibinfo{volume}{II}, \bibinfo{pages}{188--193}
  (\bibinfo{publisher}{Springer}, \bibinfo{year}{1991}).
\newblock \bibinfo{note}{Originally published in Russian in \emph{Bull. Acad.
  Sci. USSR, Phys.} Ser. 1 355, 1937}.

\bibitem{Johnson_Mehl_1939}
\bibinfo{author}{Johnson, W.~A.} \& \bibinfo{author}{Mehl, R.~F.}
\newblock \bibinfo{journal}{\bibinfo{title}{Reaction kinetics in processes of
  nucleation and growth}}.
\newblock {\emph{\JournalTitle{Trans. Amer. Inst. Min. Met. Eng.}}}
  \textbf{\bibinfo{volume}{135}}, \bibinfo{pages}{416--442}
  (\bibinfo{year}{1939}).

\bibitem{Avrami_1939}
\bibinfo{author}{Avrami, M.}
\newblock \bibinfo{journal}{\bibinfo{title}{Kinetics of phase change \mbox{I}:
  General theory}}.
\newblock {\emph{\JournalTitle{J. Chem. Phys.}}} \textbf{\bibinfo{volume}{7}},
  \bibinfo{pages}{1103--1112} (\bibinfo{year}{1939}).

\bibitem{Avrami_1940}
\bibinfo{author}{Avrami, M.}
\newblock \bibinfo{journal}{\bibinfo{title}{Kinetics of phase change \mbox{II}:
  Transformation-time relations for random distribution of nuclei}}.
\newblock {\emph{\JournalTitle{J. Chem. Phys.}}} \textbf{\bibinfo{volume}{8}},
  \bibinfo{pages}{212--224} (\bibinfo{year}{1940}).

\bibitem{Sekimoto_1986}
\bibinfo{author}{Sekimoto, K.}
\newblock \bibinfo{journal}{\bibinfo{title}{Evolution of the domain structure
  during the nucleation-and-growth process with non-conserved order
  parameter}}.
\newblock {\emph{\JournalTitle{Physica}}} \textbf{\bibinfo{volume}{135A}},
  \bibinfo{pages}{328--346} (\bibinfo{year}{1986}).

\bibitem{Onuki_2002}
\bibinfo{author}{Onuki, A.}
\newblock \emph{\bibinfo{title}{Phase Transition Dynamics}}
  (\bibinfo{publisher}{Cambridge University Press},
  \bibinfo{address}{Cambridge}, \bibinfo{year}{2002}).
\newblock \bibinfo{note}{Chap. 9}.

\bibitem{Bradley_Strenski_1989}
\bibinfo{author}{Bradley, R.~M.} \& \bibinfo{author}{Strenski, P.~N.}
\newblock \bibinfo{journal}{\bibinfo{title}{Nucleation and growth in systems
  with two stable phases}}.
\newblock {\emph{\JournalTitle{Phys. Rev. B}}} \textbf{\bibinfo{volume}{40}},
  \bibinfo{pages}{8967--8977} (\bibinfo{year}{1989}).

\bibitem{Porter_Easterling_1981}
\bibinfo{author}{Porter, D.~A.} \& \bibinfo{author}{Easterling, K.~E.}
\newblock \emph{\bibinfo{title}{Phase Transformation in Metals and Alloys}}
  (\bibinfo{publisher}{Van Nostrand Reinhold}, \bibinfo{address}{Wokingham,
  UK}, \bibinfo{year}{1981}).
\newblock \bibinfo{note}{Chap. 5}.

\bibitem{rubie1985kinetics}
\bibinfo{author}{Rubie, D.} \& \bibinfo{author}{Thompson, A.}
\newblock \bibinfo{title}{Kinetics of metamorphic reactions at elevated
  temperatures and pressures: an appraisal of available experimental data}.
\newblock In \emph{\bibinfo{booktitle}{Metamorphic Reactions}},
  \bibinfo{pages}{27--79} (\bibinfo{publisher}{Springer},
  \bibinfo{year}{1985}).

\bibitem{Cahn_1956a}
\bibinfo{author}{Cahn, J.~W.}
\newblock \bibinfo{journal}{\bibinfo{title}{The kinetics of grain boundary
  nucleated reactions}}.
\newblock {\emph{\JournalTitle{Acta Met.}}} \textbf{\bibinfo{volume}{4}},
  \bibinfo{pages}{449--459} (\bibinfo{year}{1956}).

\bibitem{Geelhood_et_al_2014}
\bibinfo{author}{Geelhood, K.~J.}, \bibinfo{author}{Luscher, W.~G.} \&
  \bibinfo{author}{Cuta, J.~M.}
\newblock \bibinfo{title}{\mbox{FRAPTRAN 1.5: A} computer code for the
  transient analysis of oxide fuel rods}.
\newblock \bibinfo{type}{Report} \bibinfo{number}{NUREG/CR-7023 (PNNL-19400),
  Vol. 1, Rev. 1}, \bibinfo{institution}{US Nuclear Regulatory Commission},
  \bibinfo{address}{Prepared by the Pacific Northwest National Laboratory
  (PNNL), Richland, WA, USA} (\bibinfo{year}{2014}).

\bibitem{Jernkvist_Massih_2019}
\bibinfo{author}{Jernkvist, L.~O.} \& \bibinfo{author}{Massih, A.~R.}
\newblock \bibinfo{title}{Improving the \mbox{FRAPTRAN} program for fuel rod
  \mbox{LOCA} analyses by novel models and assessment of recent data}.
\newblock In \emph{\bibinfo{booktitle}{Fuel Modelling in Accident Conditions}},
  IAEA-TECDOC-1889, Annex 2, \bibinfo{pages}{381--451}
  (\bibinfo{publisher}{IAEA}, \bibinfo{address}{Vienna, Austria},
  \bibinfo{year}{2019}).

\bibitem{hindle1983annealing}
\bibinfo{author}{Hindle, E.~D.}
\newblock \bibinfo{title}{Annealing studies of \mbox{Zircaloy-2} cladding at
  580-850 \mbox{$^\circ$C}}.
\newblock In \emph{\bibinfo{booktitle}{Dimensional stability and mechanical
  behaviour of irradiated metals and alloys. V. 1}}
  (\bibinfo{publisher}{British Nuclear Energy Society, London},
  \bibinfo{year}{1983}).

\end{thebibliography}
\end{document}